\documentclass[journal,article,submitpdftex,moreauthors,particles]{Definitions/mdpi} 
\firstpage{179} 
\pubvolume{7}
\issuenum{1}
\articlenumber{179}
\pubyear{2024}

\copyrightyear{2023}
\datereceived{6 Jan 2024} 
\daterevised{8 Feb 2024}
\dateaccepted{21 Feb 2024} 
\datepublished{27 Feb 2024} 
\hreflink{10.3390/particles7010010} 
\usepackage{ulem}

\Title{The impact of asymmetric dark matter on the thermal evolution of nucleonic and hyperonic compact stars}

\TitleCitation{The Impact of Asymmetric Dark Matter on the Thermal Evolution of Nucleonic and Hyperonic Compact Stars}

\Author{Edoardo Giangrandi$^{1,2}$\orcidA{}, Afonso {\'A}vila $^{1}$\orcidB{}, Violetta Sagun $^{1}$\orcidC{}, Oleksii Ivanytskyi$^{3}$\orcidD{} and Constança Providência${^1}$\orcidE{}*}

\AuthorNames{Edoardo Giangrandi, Afonso {\'A}vila, Violetta Sagun, Oleksii Ivanytskyi, Constança Providência}
\AuthorCitation{Giangrandi, E.; {\'A}vila, A.; Sagun, V.; Ivanytskyi, O.; Providência, C.}

\address{%
$^{1}$ \quad CFisUC, Department of Physics, University of Coimbra, Rua Larga P-3004-516, Coimbra, Portugal;\\
$^{2}$ \quad Institut für Physik und Astronomie, Universität Potsdam, Karl-Liebknecht-Str.24-25, Potsdam, Germany;\\
$^{3}$ \quad Incubator of Scientific Excellence---Centre for Simulations of Superdense Fluids, University of Wrocław, 50-204, Wroclaw, Poland}
\corres{Correspondence: edoardo.giangrandi@uni-potsdam.de, afonso.avila@student.uc.pt, violetta.sagun@uc.pt, oleksii.ivanytskyi@uwr.edu.pl, cp@uc.pt}

\abstract{We investigate the impact of asymmetric fermionic dark matter (DM) on the thermal evolution of neutron stars (NSs), considering a scenario where DM interacts with baryonic matter (BM) through gravity. Employing the two-fluid formalism, our analysis reveals that DM accrued within the NS core exerts an inward gravitational pull on the outer layers composed of BM. This gravitational interaction results in a noticeable increase in baryonic density within the core of the NS. Consequently, it strongly affects the star's thermal evolution by triggering the early onsets of the direct Urca (DU) processes, causing enhanced neutrino emission and rapid star cooling. Moreover, the photon emission from the star's surface is modified due to a reduction of radius. We demonstrate the effect of DM gravitational pull on nucleonic and hyperonic DU processes that become kinematically allowed even for NSs of low mass. We then discuss the significance of observing NSs at various distances from the Galactic center. Given that the DM distribution peaks toward the Galactic center, NSs within this central region are expected to harbor higher fractions of DM, potentially leading to distinct cooling behaviors.}


\begin{document}

\section{Introduction}
\label{subsec:intro}

The remarkable compactness of neutron stars (NSs) makes them the densest objects accessible by direct observations. Due to this, NSs serve as ideal laboratories to study the properties of strongly interacting matter, as well as to probe General Relativity and explore physics beyond the Standard Model~\citep{Baym:2017whm,Kramer:2021jcw}. 
Throughout the entire lifetime of an NS, a sizeable amount of dark matter (DM) could be accrued within the star's interior, affecting the matter distribution, mass, radius, etc.~\citep{Goldman:2013qla,Ivanytskyi:2019wxd, Sagun:2021oml,Karkevandi:2021ygv,Sagun:2022ezx,Giangrandi:2022wht, Das:2020ecp,Dutra:2022mxl,Diedrichs:2023trk,Lenzi:2022ypb}. 
An NS typically originates from a main sequence star of mass ranging from 8 to 20~$M_{\odot}$ that at the end of its lifetime undergoes a supernova explosion~\citep{Heger_2003}. On the other hand, this stellar journey starts from the gravitational collapse of molecular cloud regions surpassing the Jeans limit. It is worth highlighting that these proto-clouds may already harbor traces of DM, fastening up the collapse and ultimately creating newly formed stars admixed with a significant amount of DM~\citep{Yang:2020bkh}. As the star evolves, it undergoes a core-collapse supernova, during which DM may potentially be created and/or accrued in the resulting remnant, namely, an NS~\citep{Meyer:2020vzy}. Once an NS is born, it can accumulate DM particles from the surrounding galactic medium, increasing the DM fraction within the object~\citep{Brito_2015,Kouvaris:2010jy}. The latter case is valid for asymmetric DM, which refers to models characterized by an asymmetry in the number densities of particles and antiparticles. 

An estimation of the amount of DM being accreted through the Bondi accretion in the Galactic center gives up to 0.01\% of the star’s total mass~\citep{Ivanytskyi:2019wxd}. In addition, based on the predictions of many cosmological models, a rapid DM accumulation leading to a higher DM fraction inside the star could occur while passing through an extremely dense subhalo with primordial DM clumps~\citep{Bramante:2021dyx}. Thus, primordial density perturbations could result in a large fraction of DM forming gravitationally collapsed objects or clumps residing in subhalos~\citep{Erickcek:2011us,Buckley:2017ttd} or even the formation of dark compact objects~\citep{Dengler:2021qcq,Maselli:2017vfi,2015PhRvD..92f3526K}.

As DM is gravitationally trapped by an NS, it can form different configurations depending on its properties, e.g. either sinking to the center and forming a dense core, or creating a dilute halo embedding the whole NS. In the former scenario, heavy DM particles tend to create a dense inner region inside a baryonic star. The gravitational pull of the DM core leads to a denser and more compact NS configuration characterized by smaller maximum gravitational mass and radius compared to a purely baryonic NS. Consequently, these configurations can be perceived as the ones characterized by an apparent softening of the baryonic matter (BM) equation of state (EoS), or, equivalently, lower pressure at a certain baryonic density~\cite{Giangrandi:2022wht}. 

NSs with a dense DM core are more compact and harder to be deformed by an external gravitational field, a phenomenon that can be probed with forthcoming gravitational wave (GW) detections of binary NS mergers and black hole-NS mergers by assessing the tidal polarizability parameter $\Lambda$ during the inspiral phase of the merger~\citep{Giangrandi:2022wht,Sagun:2021oml,Karkevandi:2021ygv,Diedrichs:2023trk}. Moreover, the first studies show that DM might manifest itself with an additional peak or pronounced oscillation mode in the post-merger GW spectrum~\cite{Ellis:2017jgp}, generation of an exotic waveform~\cite{Giudice:2016zpa}, modification in the kilonova ejecta~\cite{Emma:2022xjs}, formation of a one-arm spiral instability~\cite{Bezares:2019jcb}, etc.~\cite{Hippert:2022snq,Ruter:2023uzc,Bauswein:2020kor}. These smoking-gun signals are expected to become feasible with the next generation of GW detectors, i.e. the Einstein Telescope~\cite{Punturo:2010zz}, Cosmic Explorer~\cite{Mills:2017urp}, and NEMO~\cite{Ackley:2020atn}. On the other hand, when the radius of the DM component surpasses the one of the BM, a halo structure ensues, enveloping the entire NS~\citep{Sagun:2022ezx}. This hints at an increase of the NS gravitational mass, mimicking a stiffening of the BM EoS~\citep{Ivanytskyi:2019wxd, Shakeri:2022dwg}. Diluted DM distributed around the NS is more susceptible to deformation, consequently impacting the tidal polarizability parameter $\Lambda$ of the star itself~\citep{Sagun:2021oml,Diedrichs:2023trk}.

Another approach to probing the presence of DM within NSs involves the study of their thermal evolution. The standard cooling of NSs includes a combination of thermal radiation emitted from the surface and neutrino emission originating in the NS interior. This offers an opportunity to relate the particle composition, the effect of nucleonic and quark superfluidity/superconductivity, as well as the other properties of matter to the NS surface temperature through X-ray observations~\cite{Ho:2014pta,Sagun:2023rzp}. Notably, thermal evolution depends on several different phenomena, occurring throughout the NS thermal history. The latter can be split into three primary stages: a newly formed NS with a thermally decoupled core and crust (age $\lesssim$ 100 years)~\cite{Sales:2020aad}, the subsequent phase dominated by neutrino emission (age 100-10$^5$ years), and a final phase dominated by photon emission (age $\gtrsim$ 10$^5$ years)~\cite{Page:2004fy,Yakovlev:2004iq}. The initial phase corresponds to the duration required for the core and crust of an NS to reach a thermal equilibrium, referred to as the thermal relaxation time. During this stage, the surface temperature of the NS remains constant as neutrinos gradually diffuse through the optically thick medium, supplying enough energy to counterbalance the cooling. As neutrinos reach the stellar surface, the star's temperature drops as a consequence of the thermal connection between the core and crust. 

In the following stage, NS cooling is primarily determined by the particle composition of the NS core, which is defined by the BM EoS. Thus, the high proton fraction and the existence of hyperons in the NS interior are especially interesting as they allow the occurrence of $\beta$-decay, and its inverse process, known as the direct Urca (DU) process~\citep{Maxwell:1986pj,Prakash:1992zng,Kaminker:2016ayg}. The DU process, once kinematically allowed by the Fermi momenta triangle inequality condition, leads to strong neutrino emission, and, consequently, to a rapid stellar cooling~\citep{Lattimer:1991ib,Page:2005fq, Potekhin:2015qsa}. In the case where the threshold condition is not met, the less efficient modified Urca (MU) process becomes 
the dominant one. This process does not have a threshold, as the presence of a spectator nucleon facilitates the reaction. The MU together with the nucleon-nucleon bremsstrahlung and the Cooper pair breaking and formation (PBF) mechanism contribute to the slow/intermediate cooling~\cite{Page:2005fq}.

At density $n=2n_0-2.5n_0$, with $n_{0}$ being the nuclear saturation density, hyperons become new degrees of freedom~\cite{Tolos:2017lgv,Li:2018qaw,Sedrakian:2022kgj}. The presence of hyperons fundamentally alters the EoS, resulting in its significant softening and the corresponding reduction in the maximum mass achievable for NSs~\citep{Fortin:2016hny,Raduta:2017wpp}. As a consequence, 
hyperonic EoS barely provide the NS masses of $~2$ M$_\odot$~\citep{Demorest:2010bx,Antoniadis:2013pzd,NANOGrav:2019jur,Yamamoto:2017wre}. 
While providing the existence of heavy NSs with masses above $\sim 2$ M$_\odot$, stiffer hadronic EoSs (see e.g. Ref.~\cite{Shahrbaf:2022upc}) are discriminated by the observational constraints on the tidal deformability~\cite{Abbott_2018}. These difficulties are naturally overcome within the scenario of early deconfinement of quark matter before the hyperon onset~\cite{Ivanytskyi:2022oxv,Ivanytskyi:2022wln,Ivanytskyi:2022bjc}. In this case, all the observational constraints can be fulfilled simultaneously~\cite{Gartlein:2023vif}.

The presence of hyperons affects the NS thermal evolution as well. As a matter of fact, hyperonic DU takes place at a very low fraction of hyperons, typically of the order of several percent. In other words, the onset of the process is at densities slightly above the point where hyperons become energetically favorable~\cite{Fortin:2020qin,Fortin:2021umb}. As it will be discussed in Sec.~\ref{sec:Thermal}, in comparison to it, the nucleonic DU is highly dependent on the proton fraction and, therefore, might be allowed only at much higher baryonic densities, depending on the density dependence of the symmetry energy. 

After $\sim 10^{5}$ years since the NS formation, the neutrino emission processes become less relevant, and photon emission from the surface starts to take the dominant role. Further on, an NS keeps on cooling down until it becomes undetectable by X-ray telescopes, and the peak of its black body radiation shifts to longer wavelengths. They cool down so drastically that they enter the sensitivity band of the James Webb Space Telescope (JWST). Thus, the JWST becomes an important tool in searching for old NSs~\cite{Baryakhtar:2017dbj,Raj:2017wrv,Chatterjee:2022dhp}. However, deviations from this cooling scenario may occur due to additional heating or cooling channels. Accretion of matter from a companion star~\citep{Wijnands:2017jsc}, magnetic field decay~\citep{Aguilera:2007dy}, rotochemical heating~\cite{Hamaguchi:2019oev} as well as decay or annihilation of DM particles to Standard model states inside the star increase its surface temperature~\citep{Berryman:2022zic,Baym_2018,Motta_2018,Motta_2018b,2010PhRvD..81l3521D}. The latter could be probed in both middle- and late-time heating scenarios~\cite{AngelesPerez-Garcia:2022qzs,Hamaguchi:2019oev}. Old NSs may exhibit a plateau in their cooling curves, corresponding to a balance between the energy loss and heating caused by DM self-annihilation~\cite{Kouvaris_2008,Bramante:2023djs}. The late-time heating is another sensitive probe of DM as the influence of nuclear superfluidity/superconductivity, magnetic field, and other factors that may wash out the effect, become irrelevant at this age. Accumulated asymmetric DM also contributes to a star's heating by depositing the kinetic energy gained during the infall, so-called ``dark kinetic heating''~\cite{Baryakhtar:2017dbj,Raj:2017wrv}. The primary contribution to the cooling of old NSs is photon emission from the surface, as neutrino cooling is suppressed, making it feasible to probe the heating effects of DM. The ongoing, i.e. XMM-Newton and Chandra~\cite{Haberl:2003cg}, NICER~\cite{Riley:2021pdl, Miller:2021qha}, and future X-ray surveys, i.e. ATHENA~\cite{Cassano:2018zwm}, eXTP~\cite{eXTP:2018kws}, and STROBE-X~\cite{STROBE-XScienceWorkingGroup:2019cyd}, are expected to increase the number of mass, radius, and surface temperature determinations, including the old NS detections, shedding light on the possible DM impact on their thermal evolution.

On the other hand, light DM particles emitted from an NS carry away energy, further cooling down the star~\cite{Kumar:2022amh}. In comparison to the heavy DM particles with mass $\ge$MeV that could be evaporated from the NS's surface~\citep{Garani:2021feo}, light particles like axions could be emitted through the star. Thus, various studies have explored the effects of axion emission on NS and proto-NS thermal evolution~\cite{Dietrich:2019shr}, where axions produced within the NS core in nucleon bremsstrahlung or PBF mechanism, will contribute to the star's temperature drop~\cite{Sedrakian:2015krq,Buschmann:2019pfp, Buschmann:2021juv}. 

This study focuses on asymmetric DM consistent with the
$\Lambda$CDM model predicting cold (non-relativistic) collisionless DM particles~\cite{Planck:2018vyg}. Accumulated asymmetric DM, which interacts with BM only through gravity, is considered to have no impact on the thermal evolution of NSs as it does not annihilate, is heavy, and does not participate in any processes involving BM. However, this study shows that accrued DM by modifying the BM distribution and compactness of the star, affects the DU processes responsible for rapid cooling. As a result, it is triggered in stars of lower mass. In this work, we extend our previous analysis~\citep{Avila:2023rzj} by considering hyperons. We investigate an interplay between hyperons and DM in the inner NS region and the modification of the stars' thermal evolution. This analysis assumes that a DM fraction remains unchanged or accretion is negligibly small after the NS formation, and, therefore, dark kinetic heating does not affect the cooling.

The paper is organized as follows. In Section \ref{sec:EoS}, we present the considered EoSs for BM and DM. In Section \ref{sec:2fluid} we discuss the two-fluid formalism, while
Section \ref{sec:Thermal} reviews the NS thermal evolution and all the nucleonic processes involved in it. In Section \ref{sec:Results} we provide the results of the cooling simulations for different DM fractions. Conclusions are presented in Section \ref{sec:Concl}. Throughout the article, we use the unit system in which $\hbar=c=G=1$.

\section{Baryonic and dark matter EoSs}
\label{sec:EoS}

\subsection{Baryonic matter}
\label{subsec:BarMatter}

To account for the BM uncertainties at high densities, we adopt three different models characterized by different particle compositions, stiffness, and nuclear matter properties at the saturation density. Specifically, we have chosen these models based on the different values of their symmetry energy slope $L$ and the incompressibility factor $K_{0}=9(\frac{\partial p~}{\partial n_\mathrm{B}})_{n_0}$, at the normal nuclear density. The first of these models is the Induced Surface Tension (IST) EoS, which is formulated considering the hard-core repulsion among particles. Notably,~\citet{Sagun:2013moa} demonstrated that in a dense medium, a short-range repulsion between particles gives rise to an additional contribution to the single-particle energy, the IST term~\cite{Bugaev:2021pwn}. At high densities, the IST contribution vanishes leading to a correct transition from the excluded volume approach to the proper volume regime. In the limit of a dilute gas, the IST EoS accurately recovers the first four virial coefficients of hard spheres. The values of the hard-core radius of nucleons were obtained from the fit of the particle ratios created in heavy-ion collision experiments in a wide range of the center of mass collision energies~\cite{Sagun:2017eye, Bugaev:2021pwn}.
Furthermore, the IST EoS reproduces the nuclear liquid-gas phase transition, including its critical endpoint~\cite{Sagun:2016nlv}. The generalization of the model to account for the long-term attraction and asymmetry energy was formulated in agreement with the NSs observations and tidal deformability from the GW170817 binary NS merger ~\cite{Sagun:2018cpi}. In this work, we utilize Set B of the IST EoS developed in~\citep{NSOscillationsEoS}. 

The second considered model is the nucleonic relativistic mean-field FSU2R EoS~\cite{Tolos:2017lgv}, which is the generalization of the FSU2 EoS~\cite{Chen:2014sca}. The FSU2R model is characterized by a smaller symmetry energy and neutron matter pressure, allowing it to describe NSs with smaller radii compared to the original FSU2 EoS. The model parameters were determined by fitting the binding energies, charge radii, and monopole response of atomic nuclei across the periodic table. Importantly, the FSU2R model reproduces equally well the properties of nuclear matter and finite nuclei. 

To study the effects of hyperons on the NS thermal evolution, we adopt the FSU2H EoS in which the parameters of the isospin-symmetric matter were adjusted to solve the hyperon puzzle, providing a stiffer EoS and fulfilling the heavy NS constraints~\cite{Tolos:2017lgv}. The adjustments were made only above two saturation densities, i.e. after hyperonic onset, to preserve the properties of the model at saturation density and to fulfill finite nuclei and stellar radii. The model is parameterized with the values of the hyperons couplings and potentials ($\Lambda$, $\Sigma$, and $\Xi$) as described by~\citet{Fortin:2021umb}. However, following Ref.~\citep{Gal:2016boi}, the $\Sigma$ potential $U^{(N)}_\Sigma$ was set at $30$ MeV and not spanned in a wide range of values. As a result, this parametrization does not allow $\Xi^0$ hyperons to appear. Table.~\ref{tab1} summarizes the main model parameters.

\begin{table*}[ht]
\centering
\resizebox{0.85\columnwidth}{!}{%
\begin{tabular}{|c|c|c|c|c|c|c|c|}
\hline
\multirow{2}{*}{} & $n_0$                & $E/A$              & $K_0$            & $E_{sym}$        & $L$                & $M_{max}$            & $R_{1.4}$       \\
                  & $\mathrm{[fm^{-3}]}$ & $\mathrm{[MeV]}$ & $\mathrm{[MeV]}$ & $\mathrm{[MeV]}$ & $\mathrm{[MeV]}$ & $\mathrm{[M_\odot]}$ & $\mathrm{[km]}$ \\ \hline
IST EoS           & 0.16                 & -16.00           & 201.0            & 30.0             & 93.19            & 2.084                & 11.4            \\ \hline
FSU2R EoS         & 0.1505               & -16.28           & 238.0            & 30.7             & 46.90            & 2.048                & 12.8            \\ \hline
FSU2H EoS         & 0.1505               & -16.28           & 238.0            & 30.5             & 44.50            & 1.992                & 12.7            \\ \hline
\end{tabular}%
}
\caption{Parameters of the IST, FSU2R and FSU2H models. The table includes the saturation density $n_{0}$, energy per baryon $E/A$, incompressibility factor $K_{0}$, symmetry energy $E_{sym}$, and its slope $L$ at saturation density, as well as the maximum gravitational mass $M_{max}$, and radius of the 1.4 M$_{\odot}$ star.}
\label{tab1}
\end{table*} 

\begin{figure}[htp]
    \centering
    \includegraphics[width=0.45\columnwidth]{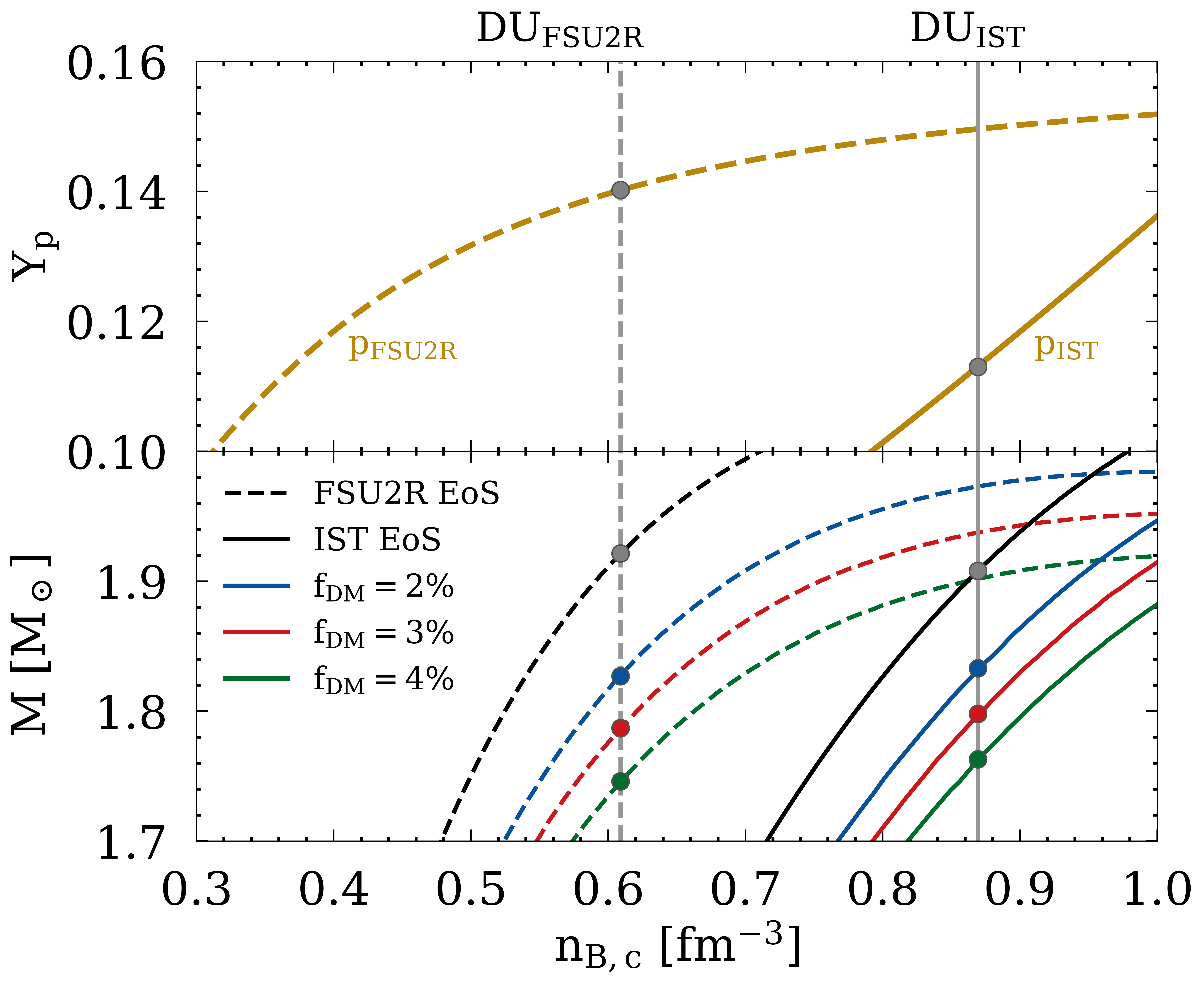}
    \includegraphics[width=0.45\columnwidth]{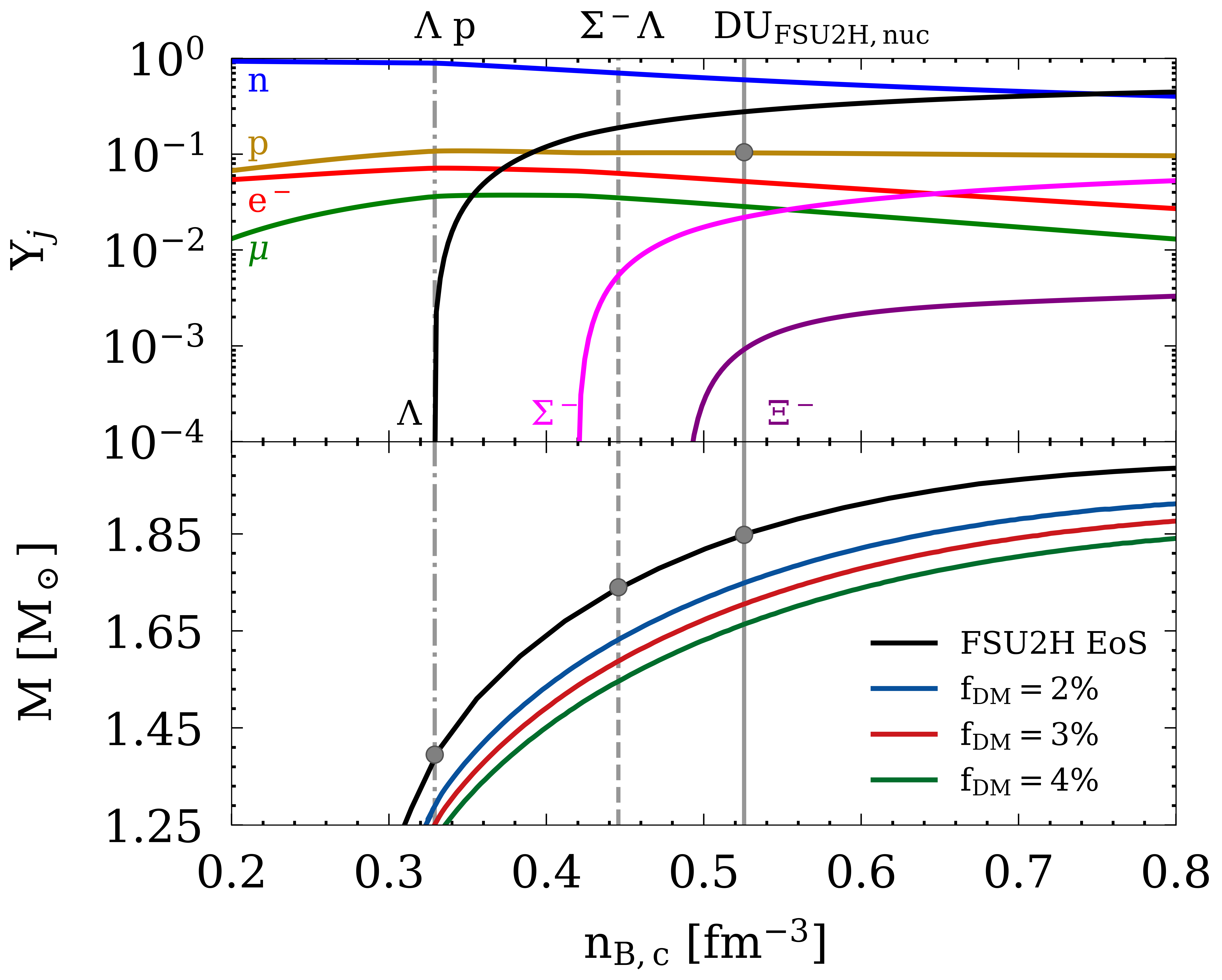}
    \caption{{\bf Left panel:} The proton fraction, $Y_p$, (top) and total gravitational mass of the DM-admixed and baryonic NSs (bottom) as a function of the central baryonic density, $n_{B,c}$, for the IST and FSU2R EoSs. On the bottom panel, the royal blue, red, and green curves represent the relative DM fractions equal to 2\%, 3\%, and 4\%, respectively. The vertical solid and dashed gray lines correspond to the central BM densities of the stars at which the DU process is activated for the IST and FSU2R EoSs, respectively. The intersection points depict the nucleonic DU threshold. 
    {\bf Right panel:} The particle fractions, $Y_j$, (top) and total gravitational mass of the DM-admixed and baryonic NSs (bottom) as a function of the central baryonic density, $n_{B,c}$, for the FSU2H EoS. The vertical solid, dashed, and dashed-dotted gray lines correspond to the central BM densities at which the DU processes are activated, i.e. nucleonic ($np$) and hyperonic ($\Lambda p$ and $\Sigma^-\Lambda$), respectively. 
    }
    \label{fig:DU_ParticleFraction}
\end{figure}

The upper part of the left panel of Fig.~\ref{fig:DU_ParticleFraction} shows the relative fraction of proton for the IST and FSU2R EoS that have only $n, p,\ e^-$ as degrees of freedom. For comparison, the FSU2H EoS includes also $\mu$, $\Lambda$, $\Xi^-$, and $\Sigma^-$ hyperons. All particle fractions of the FSU2H EoS are depicted on the upper part of the right panel of Fig.~\ref{fig:DU_ParticleFraction}. These particles appear as a result of the attractive nature of the $\Lambda N$ and $\Xi N$ interactions~\citep{Fortin:2016hny,Li:2018jvz}.

Another important effect of hyperons on the matter properties is related to modifications of the particle fractions upon their onset. As shown on the upper right panel of Fig.~\ref{fig:DU_ParticleFraction}, as $\Lambda$ hyperons start to appear, their decay into protons leads to an increase of the latter ones.

All the BM EoSs were supplemented by the Haensel-Zdunik (HZ) EoS for the outer crust and the Negele-Vautherin (NV) EoS for the inner crust~\citep{1990A&A...227..431H,Negele:1971vb}.

\subsection{Dark matter}
\label{subsec:DarMatter}
Hereby, DM is described as a fermionic gas of relativistic particles with spin 1/2~\citep{Nelson:2018xtr,Ivanytskyi:2019wxd,Sagun:2021oml}. 
This model is the simplest realization of the asymmetric DM that due to the Pauli blocking principle can resist a gravitational collapse and form stable configurations inside NSs~\citep{Bell:2013xk}. Thus, DM particles are characterized by a small number of parameters, i.e. mass $m_{DM}$ and degeneracy factor $g_{DM}=2$, making it a widely used model~\citep{Ivanytskyi:2019wxd}.

The pressure $p_{DM}$ and energy density $\varepsilon_{DM}$ of the system can be written down as follows
\begin{equation}
\begin{cases}
    p_{DM} = \frac{g_{DM}}{48\pi^2}[\mu_{DM} k_{DM} (2\mu_{DM}^2-5m_{DM}^2)+3m_{DM}^4\ln(\frac{\mu_{DM}+k_{DM}}{m_{DM}})],\\
    \varepsilon_{DM}=\mu_{DM} n_{DM} -p_{DM},
\end{cases}
\end{equation}
where $\mu_{DM}$, $n_{DM}=\partial p_{DM} / \partial \mu_{DM}$ and $k_{DM}=\sqrt{\mu_{DM}^2-m_{DM}^2}\theta(\mu_{DM}-m_{DM})$ the DM  chemical potential, number density, and Fermi momentum, respectively.

In this work, we consider the DM particle mass $m_{DM}$=1 GeV. This choice of the DM particle's mass is motivated by the intriguing similarity with nucleons. A study of the broad particle mass range of the fermionic DM in the MeV-GeV scale can be found in Ref.~\cite{Ivanytskyi:2019wxd}.


\section{Two-fluid formalism}
\label{sec:2fluid}

Observations of the merging clusters of galaxies, i.e. the Bullet Cluster (1E 0657-56), Pandora's Cluster (Abell 2744), MACS J0025.4-1222~\citep{Eckert:2022qia, Clowe_2006, Randall:2008ppe}, and DM direct detection experiments~\citep{Goodman:2010ku} show that the BM-DM cross-section is many orders of magnitude lower than the typical nuclear one, $\sigma_{DM}\sim 10^{-45}\ \mathrm{cm}^2\ll \sigma_N\sim10^{-24}\ \mathrm{cm}^2$. Hence, with appropriate accuracy, we can assume no interactions between BM and DM except the gravitational one. 

The Einstein's field equations can be written down as
\begin{equation}
    R^{\mu\nu}+\frac{1}{2}g^{\mu\nu}R = 8\pi T^{\mu\nu}.
\end{equation}
The stress-energy tensor has a form
\begin{equation}
    T^{\mu\nu}=[\varepsilon+p]u^\mu u^\nu + p g^{\mu\nu},
\end{equation}
where $\varepsilon, p$, and $u^\mu$ are the proper energy density, pressure, and four-velocity of the fluid component. Due to a negligibly weak nongravitational interaction between BM and DM the only nonvanishing components in the perfect fluid approximation in the stress-energy tensor are
\begin{equation}
T^{11}=T^{22}=T^{33}=p_{BM}+p_{DM},~~~~~~T^{44}=\varepsilon_{BM}+\varepsilon_{DM}.
\end{equation}
Consequently, each fluid individually satisfies the equations of motion of a single fluid, hence the energy-momentum conservation, $\nabla^\mu T_{\mu\nu}^i=0$. Thus, the Tolmann-Oppenheimer-Volkoff (TOV) equation~\citep{PhysRev.55.364,PhysRev.55.374} that describes a non-rotating spherically symmetric star in hydrostatic equilibrium can be split into two coupled first-order ODEs, one for each component~\citep{Ivanytskyi:2019wxd}.
\begin{equation}
\label{TOV}
\frac{dp_i}{dr}=-\frac{(\varepsilon_i +p_i)(M_\mathrm{tot}+4\pi r^3p_\mathrm{tot})}{r^2\left(1-{2M_\mathrm{tot}}/{r}\right)}.
\end{equation}
Here $p_\mathrm{tot}\equiv p_{BM}+p_{DM}$ and $M_\mathrm{tot} = M_{BM}(R_{BM})+M_{DM}(R_{DM})$ are the total pressure and total gravitational mass enclosed in a sphere of radius $r$ which can be written as
\begin{equation}
\label{Mas}
    M_i(r) = 4\pi\int^r_0 \varepsilon_i (r^\prime)r^{\prime 2}dr^\prime.
\end{equation}
The radii $R_i$ are evaluated using the zero-pressure condition at the surface $p_i(R_i)=0$. For a given EoS and the boundary conditions $m_i(0)=0$, $p_i(0)=p_i^c$ with $p_i^c$ being the pressure of each of the components in the center of a star we can obtain the total gravitational mass of the system, radius profile, etc. 
At the same time, the pressure of each of the components is fixed by the corresponding chemical potential.
Therefore, in this work, we used the boundary conditions for the chemical potentials $\mu_i(0)=\mu_i^c$ with $\mu_i^c$ being the corresponding central values.
In this case, the system of Eqs.~\eqref{TOV} can be converted to
\begin{equation}
\label{eq:TOV2}
    \frac{d \ln \mu_BM}{dr}=\frac{d \ln \mu_{DM}}{dr} = -\frac{M_\mathrm{tot}+4\pi r^3 p_\mathrm{tot}}{r^2(1-2M_\mathrm{tot}/r)},
\end{equation}
obtained in Ref.~\cite{Ivanytskyi:2019wxd}.It follows from these relations that the chemical potentials of BM and DM scale proportionally according to their central values.

\begin{figure}[t]
    \centering
    \includegraphics[scale=0.4]{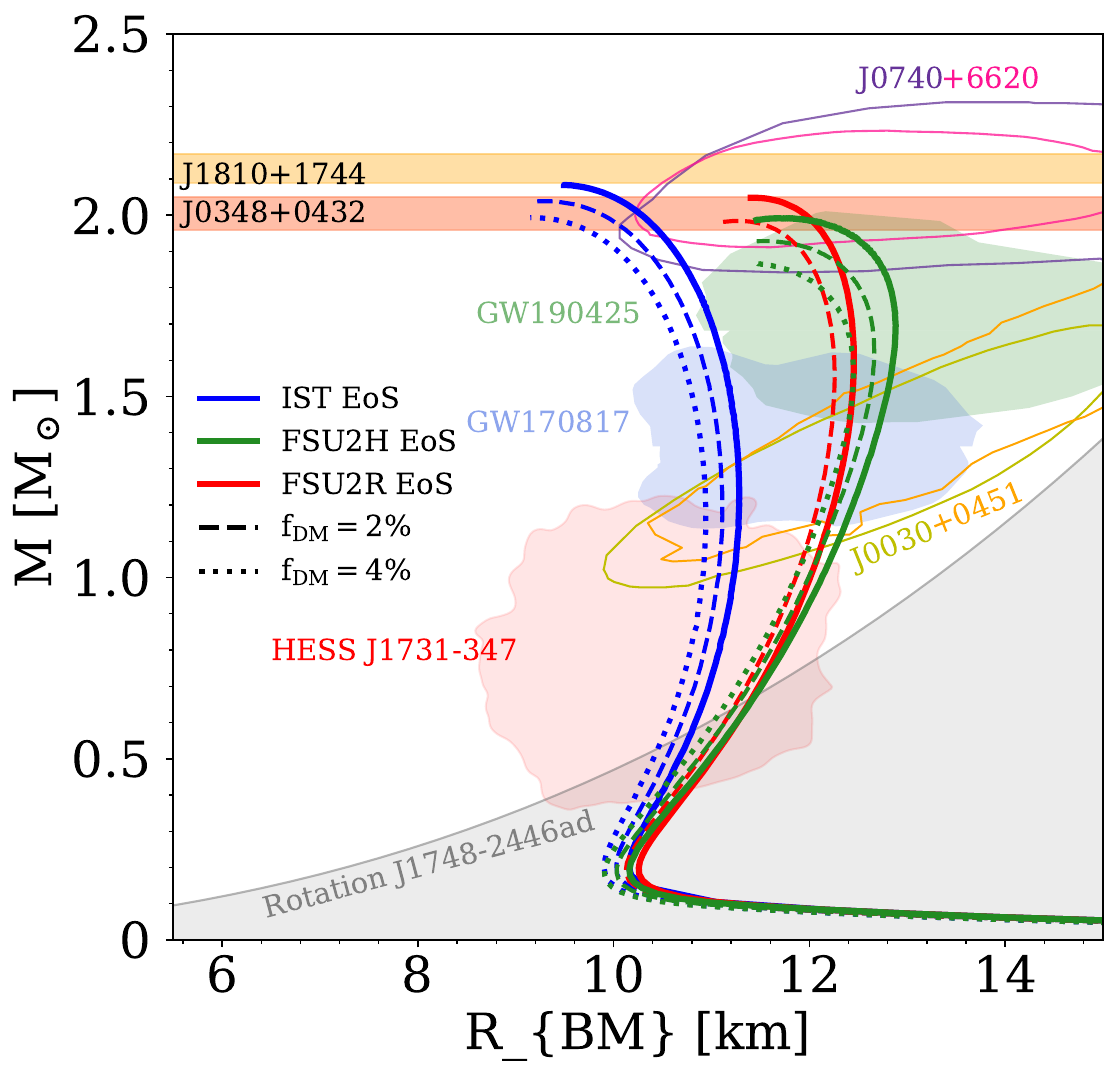}
    \caption{Total gravitational mass of DM-admixed and baryonic NSs as a function of their baryonic radius $R_B$ calculated for the DM particle mass $m_{DM}$=1 GeV. Solid curves correspond to pure BM stars described by the IST (blue curve), FSU2R (red curve), and FSU2H (green curve) EoSs. Dashed and dotted curves depict the M-R relations obtained for the relative DM fractions equal to 2\% and 4\%, respectively. Orange and dark yellow bands represent $1\sigma$ constraints on the mass of PSR J0348+0432~\citep{Antoniadis:2013pzd} and PSR J1810+1744~\citep{Romani:2021xmb}. While olive green and light orange contours show the NICER measurements of PSR J0030+0451~\citep{Miller:2019cac, Riley:2019yda}, purple and magenta contours correspond to the PSR J0740+6620 measurement~\citep{Raaijmakers:2021uju, Miller:2021qha}. Observations of GW170817~\citep{LIGOScientific:2018cki} and GW190425~\citep{LIGOScientific:2020aai} binary NS mergers by LIGO-Virgo collaboration are shown in blue and green. The $2\sigma$ contour of HESS J1731-347~\citep{Doroshenko2022} is plotted in light red. The shaded gray region is excluded by the rotation of the fastest spinning pulsar PSR J1748-2446ad~\citep{Hessels:2006ze}.}
    \label{fig:MRcurves}
\end{figure}

It is convenient to define the DM mass fraction as
\begin{equation}
\label{eq:DMFrac}
    f_\mathrm{DM} = \frac{M_\mathrm{DM}}{M_\mathrm{tot}}.
\end{equation}
The mass-radius curves for the aforementioned hadronic models are represented in Fig.~\ref{fig:MRcurves}, including the DM-admixed configurations. They were obtained by solving the TOV Eqs.~\eqref{eq:TOV2} at varying the central value of the BM chemical potential and adjusting the central value of the DM one so that its fraction is kept equal to the values indicated in the figure. 

Fig.~\ref{fig:MRcurves} shows the solid M-R curves for baryonic stars described by the IST, FSU2R, and FSU2H EoSs in blue, red, and green, respectively. The dashed and dotted curves represent the DM-admixed configurations with DM fractions of 2\% and 4\% for the corresponding BM EoSs. 

\section{NS thermal evolution}
\label{sec:Thermal}

The thermal evolution of isolated NSs is determined by the energy balance equation which can be written down as
\begin{equation}
C_{\mathrm{v}} \frac{d T{_s^\infty}}{d t}=-L_\nu^\infty-L_\gamma^\infty \pm H^\infty.
\label{Eq:4}
\end{equation}
Here $C_\mathrm{v}$ is the total specific heat of the stellar matter, $T{_s^\infty}$, $L_\nu^\infty$, $L_\gamma^\infty$ are the red-shifted surface temperature, neutrino luminosity, and photon luminosity, respectively~\citep{Page:2005fq}. The last term on the right-hand side of Eq.~\eqref{Eq:4} refers to extra mechanisms contributing to heating ($+$) or cooling ($-$) the compact star. In this work, no additional cooling/heating mechanisms were considered, hence $H^\infty \equiv 0$. As photons are emitted from the stellar surface, the photon luminosity $L_\gamma = 4\pi \mathrm{R_{B}^2} \sigma T{_s}^4$ strongly depends on the stellar baryonic radius $R_{B}$. Here, $\sigma$ is the Stefan-Boltzmann constant. Once the photon luminosity in the local frame is obtained, it can be red-shifted to obtain the corresponding value in the frame of an observer placed at infinity $L_\gamma^\infty$. This operation can be performed by multiplying the quantity by the factor $e^{\Phi}$, i.e. by the $g_{tt}$ component of the metric, derived by integrating the following ODE
\begin{equation}
\frac{d\Phi}{dr}= -\frac{dp_{\mathrm{tot}}}{dr}\frac{1}{\varepsilon_\mathrm{tot}+p_\mathrm{tot}}.
\label{Eq:5}
\end{equation}
Here $\varepsilon_\mathrm{tot}=\varepsilon_{BM}+\varepsilon_{DM}$ is the total energy density. 

On the other hand, neutrinos are emitted throughout the star as the NS matter is optically thin for these particles. The neutrino luminosity is highly sensitive to various factors, such as Cooper pairing between particles, particle composition, their fractions, etc. The latter ones highly depend on the underlying BM EoS~\citep{Lattimer:1991ib,Page:2005fq, Potekhin:2015qsa}. In fact, the enhanced cooling due to the neutron $\beta$-decay and its inverse process, known as the DU process,
\begin{equation}
\label{DUnucl}
    \begin{cases}
        n  \rightarrow p+\ell+\overline{\nu}_\ell \\
        p  + \ell \rightarrow n + \nu_\ell
    \end{cases}
\end{equation}
occur when the Fermi momenta of the involved particles satisfy the triangle inequality condition, $p_{F,i}+p_{F,j} \ge p_{F,k}$, where $i$, $j$ and $k$ are the involved particles and their cyclic permutations. In Eq.~\eqref{DUnucl} $\ell$ stands for a lepton, $e^{-}$ or $\mu$. The triangle inequality condition ensures that, in the presence of strongly degenerate fermions, it can only take place when the energies of the involved particles are close to their Fermi energies. In the case of proton, electron, and neutron composition the triangle condition, reads as $p_{Fp}+p_{Fe}\ge p_{Fn}$, with $p_{Fp}$, $p_{Fe}$, and $p_{Fn}$ being the Fermi momenta of particles. Taking into account charge neutrality and the connection between Fermi momenta and the number density of each particle, the proton fraction ($Y_{p}$) must be above approximately 11\%~\citep{Lattimer:1991ib}. The latter value is obtained for n,p,$e^{-}$ matter. While muons are present, the minimal proton fraction attains a larger value
\begin{equation}
Y_{p}=\frac{1}{1+(1+x_{e}^{1/3})^{3}},
\label{minfrac_nucl}
\end{equation}
where $x_{e}=n_{e}/(n_{e} + n_{\mu})$ and $n_{e}$, $n_{\mu}$ are densities of electrons and muons.

The triangle condition might not be met for some models, meaning that up to the TOV mass, the nucleonic DU process is not triggered in the stellar core. In our study, all three considered models, the IST, FSU2R, and FSU2H, have the onset of the DU at the central densities and corresponding masses shown in Tables~\ref{Table:DUrcas_IST_FSU2R},~\ref{Table:DUrcas_FSU2H}. Further on, we will refer to $n_{DU}$ and mass $M_{DU}$ as the baryonic density at which the DU process onsets and the corresponding mass of the star with the central baryonic density equal to $n_{DU}$.

\begin{table}[ht]
\centering
\resizebox{0.75\columnwidth}{!}{%
\begin{tabular}{|cc|cc||cc|cc|}
\hline
\multicolumn{2}{|c|}{\multirow{3}{*}{IST EoS}}                     & \multicolumn{1}{c|}{n$_{DU}$}               & M$_{DU}$  &\multicolumn{2}{c}{\multirow{3}{*}{FSU2R EoS}}   &\multicolumn{1}{|c|}{n$_{DU}$}               & M$_{DU}$         \\  
\multicolumn{2}{|c|}{}                                & \multicolumn{1}{c|}{[$\mathrm{fm^{-3}}$]}   & [$\mathrm{M_\odot}$] &
\multicolumn{2}{|c|}{}                                & \multicolumn{1}{c|}{[$\mathrm{fm^{-3}}$]}   & [$\mathrm{M_\odot}$] \\  \hline
\multicolumn{1}{|c|}{\multirow{4}{*}{$\mathrm{f_{DM}}$}} & 0\% & \multicolumn{1}{c|}{\multirow{4}{*}{0.869}} & 1.908  &  \multicolumn{1}{c|}{\multirow{4}{*}{$\mathrm{f_{DM}}$}} & 0\% & \multicolumn{1}{c|}{\multirow{4}{*}{0.608}} & 1.921                 \\ \cline{2-2} \cline{4-4} \cline{6-6} \cline{8-8} 
\multicolumn{1}{|c|}{}                          & 2\% & \multicolumn{1}{c|}{}                       & 1.83                 & \multicolumn{1}{c|}{}                       & 2\%                 & \multicolumn{1}{c|}{}                       & 1.83                 \\ \cline{2-2} \cline{4-4} \cline{6-6} \cline{8-8} 
\multicolumn{1}{|c|}{}                          & 3\% & \multicolumn{1}{c|}{}                       & 1.80                & \multicolumn{1}{c|}{}                       & 3\%                & \multicolumn{1}{c|}{}                       & 1.79                 \\ \cline{2-2} \cline{4-4} \cline{6-6} \cline{8-8} 
\multicolumn{1}{|c|}{}                          & 4\% & \multicolumn{1}{c|}{}                       & 1.76                 & \multicolumn{1}{c|}{}                       & 4\%                 & \multicolumn{1}{c|}{}                       & 1.75                 \\ \hline
\end{tabular}%
}
\caption{Baryonic densities and total gravitational masses at the onset of the nucleonic DU process for the IST and FSU2R EoSs.}
\label{Table:DUrcas_IST_FSU2R}
\end{table}

\begin{table}[ht]
\centering
\resizebox{0.75\columnwidth}{!}{%
\begin{tabular}{|cc|cc|cc|cc|}
\hline
\multicolumn{2}{|c|}{\multirow{3}{*}{FSU2H EoS}}      & \multicolumn{2}{c|}{$\Lambda p$}                                   & \multicolumn{2}{c|}{$\Sigma^- \Lambda$}                            & \multicolumn{2}{c|}{$np$}                                          \\ \cline{3-8} 
\multicolumn{2}{|c|}{}                                & \multicolumn{1}{c|}{n$_{DU}$}               & M$_{DU}$             & \multicolumn{1}{c|}{n$_{DU}$}               & M$_{DU}$             & \multicolumn{1}{c|}{n$_{DU}$}               & M$_{DU}$             \\ 
\multicolumn{2}{|c|}{}                                & \multicolumn{1}{c|}{[$\mathrm{fm^{-3}}$]}   & [$\mathrm{M_\odot}$] & \multicolumn{1}{c|}{[$\mathrm{fm^{-3}}$]}   & [$\mathrm{M_\odot}$] & \multicolumn{1}{c|}{[$\mathrm{fm^{-3}}$]}   & [$\mathrm{M_\odot}$] \\ \hline
\multicolumn{1}{|c|}{\multirow{4}{*}{$f_{DM}$}} & 0\% & \multicolumn{1}{c|}{\multirow{4}{*}{0.332}} & 1.40                 & \multicolumn{1}{c|}{\multirow{4}{*}{0.446}} & 1.74                 & \multicolumn{1}{c|}{\multirow{4}{*}{0.534}} & 1.85                 \\ \cline{2-2} \cline{4-4} \cline{6-6} \cline{8-8} 
\multicolumn{1}{|c|}{}                          & 2\% & \multicolumn{1}{c|}{}                       & 1.32                 & \multicolumn{1}{c|}{}                       & 1.63                 & \multicolumn{1}{c|}{}                       & 1.75                 \\ \cline{2-2} \cline{4-4} \cline{6-6} \cline{8-8} 
\multicolumn{1}{|c|}{}                          & 3\% & \multicolumn{1}{c|}{}                       & 1.28                 & \multicolumn{1}{c|}{}                       & 1.59                 & \multicolumn{1}{c|}{}                       & 1.71                 \\ \cline{2-2} \cline{4-4} \cline{6-6} \cline{8-8} 
\multicolumn{1}{|c|}{}                          & 4\% & \multicolumn{1}{c|}{}                       & 1.24                 & \multicolumn{1}{c|}{}                       & 1.54                 & \multicolumn{1}{c|}{}                       & 1.66                 \\ \hline
\end{tabular}%
}
\caption{The same as in Table~\ref{Table:DUrcas_IST_FSU2R}, but for the FSU2H model. The columns correspond to different hyperonic and nucleonic DU onsets.}
\label{Table:DUrcas_FSU2H}
\end{table}

On the other hand, after hyperonization of matter in the inner core, the neutrino luminosity is amplified, triggered by the hyperonic DU~\citep{Yakovlev:2000jp}. Each of the hyperonic DU channels is open after the appearance of all the species involved in the process and the fulfillment of the momentum conservation condition~\citep{Schaab:1998zq,Prakash:1992zng}. Thus, among all hyperonic DU processes, only the following are allowed by the FSU2H EoS
\begin{equation}
    \begin{cases}
        \Lambda \rightarrow p+\ell +\overline{\nu}_\ell\\
        p+\ell \rightarrow \Lambda+ \nu_\ell
    \end{cases}
    \begin{cases}
        \Sigma^- \rightarrow \Lambda+\ell+\overline{\nu}_\ell\\
        \Lambda+\ell \rightarrow  \Sigma^- + \nu_\ell.
    \end{cases}
\end{equation}
Further on in the text, we will refer to these processes as $np$, $\Lambda p$, and $\Sigma^- \Lambda$. The remaining processes and their inverse, i.e. 
\begin{eqnarray}
        \Sigma^- &\rightarrow& \Sigma^0+\ell+\overline{\nu_\ell}\\
        \Xi^- &\rightarrow& \Xi^0 +\ell+\overline{\nu_\ell}\\
        \Xi^0 &\rightarrow& \Sigma^+ +\ell+\overline{\nu_\ell}\\
        \Xi^- &\rightarrow& \Sigma^0 +\ell+\overline{\nu_\ell}\\
        \Xi^- &\rightarrow& \Lambda +\ell+\overline{\nu_\ell}\\
        \Sigma^- &\rightarrow& n+\ell+\overline{\nu_\ell}
\end{eqnarray}
are not kinematically allowed. After the onset of hyperons, the minimum proton fraction for nucleonic DU showed in Eq.~\eqref{minfrac_nucl} changes to 
\begin{equation}
\frac{n_{p}}{n_{p}+n_{n}}= \frac{1}{1+(1+{x_{e}^{Y}}^{\frac{1}{3}})^{3}},
\label{minfrac_nucl_withHyperons}
\end{equation}
where $x^Y_{e}=\frac{n_{e}}{(n_{e} + n_{\mu} - n_{Y})}$, $n_Y = -n_{\Sigma^-}+n_{\Sigma^+}-n_{\Xi^-}$, and $n_{j}$ ($j=e,\mu,n,p,\Sigma^-, \Sigma^+, \Xi^-$) are the densities of the corresponding sort of particles~\citep{Providencia:2018ywl}. As the isospin of $\Sigma^-$ hyperon is one, its occurrence is strongly affected by the density dependence of the symmetry energy~\cite{Providencia:2018ywl}. The DU threshold densities and the star's masses described by the FSU2H model are presented in Table~\ref{Table:DUrcas_FSU2H}. 

The triggered nucleonic and hyperonic DU processes have the emissivity of $\varepsilon \sim T^{6}$ leading to an enhanced star cooling. When the threshold conditions are not met, the less efficient MU, i.e. $N+n \rightarrow N+p+\ell+\overline{\nu}_\ell$, $N+ p  + \ell \rightarrow N+n + \nu_\ell$ (with $N=n,p$,  $\ell=e^{-},\mu$), and nucleon bremsstrahlung processes become the predominant mechanisms for neutrino emission with $\varepsilon \sim T^{8}$~\citep{Iwamoto:1980eb, Iwamoto:1982zz}.

Finally, at lower temperatures, the Cooper pairing of nucleons contributes to the neutrino emissivity~\citep{Potekhin:2015qsa}. The existence of an attractive force among particles results in the formation of pairs, where the particle excitations are gapped. Thus, when the temperature falls below the critical value for neutron superfluidity and proton superconductivity, denoted as $T \ll T_{c}$, the emission of neutrinos is drastically suppressed, following the Boltzmann factor of $e^{-\Delta/T}$, where $\Delta$ corresponds to the energy gap. Once activated at the critical temperature, the effect of the PBF mechanism leads to the neutrino emissivity of the order of $\varepsilon \sim T^7$~\cite{Page:2005fq}. In the inner crust, neutrons are expected to create pairs in the singlet $^1S_0$ state and the triplet $^3P_2$ state in the core. Cooper pairs of protons occur in the singlet $^1S_0$ state in the NS core. We examine the above-mentioned nucleon pairing in this article. For simplicity, in this work, no hyperon superfluidity was considered. 
Based on the previous studies, e.g. in Ref.~\citep{Raduta:2017wpp}, we expect that incorporation of the hyperonic superfluidity would alter the onset of the hyperonic DU processes towards higher densities. Thus, the process would not be triggered right after the onset of hyperons. This could have a significant impact on the star's cooling whereas nucleonic DU is only permitted at densities exceeding the onset of hyperons, as observed in the three models studied here. However, another interplay between the nucleonic and hyperonic DU is discussed in~\citep{Raduta:2017wpp}.

To model the thermal evolution of NSs we adopted the public code \texttt{NSCool}~\citep{Neutronstarcoolingcode}, using the implementation method described by~\citet{Avila:2023rzj}. The specific heat is calculated as the sum of the contributions from its constituent particles: neutrons, protons, electrons, and hyperons (if present) using the standard \texttt{NSCool} implementation. The impact of nucleon superfluidity is accounted for by introducing control functions $R_c$ which depend on the corresponding critical temperature~\citep{Yakovlev:1999sk,Baiko:2001cj}.

To account for the DM effects, all the NS profiles were obtained using the two-fluid formalism. Despite accumulated DM does not interact with BM, it leads to BM redistribution. 
Consequently, the DM affects not only the profiles of such quantities as the total pressure, energy density, and gravitational mass, which explicitly include a DM term, but also the profiles of the metric functions, baryon density, and particle fractions. At the same time, the quantities, that do not have a DM contribution, can be directly used as an input for the single-fluid cooling formalism of the \texttt{NSCool}.
This allows us to treat a two-fluid cooling problem within the single-fluid framework.


\section{Results}
\label{sec:Results}

\subsection{Hyperonic and nucleonic DU onsets}

We start our analysis by studying three different mass configurations representing the low-, middle-, and heavy-mass stars (see Table~\ref{Table:radii}). Purely baryonic stars of 1.9 M$_\odot$ modeled by the IST and FSU2R EoSs do not exhibit a rapid cooling due to the DU process as it is triggered in heavier stars where the central density is higher (see Table~\ref{Table:DUrcas_IST_FSU2R} for details). However, heavy DM particles of $\gtrsim$ GeV scale tend to form a dense core inside a star, pulling BM inwards from the outer layers, and leading to BM redistribution. As a consequence, the baryonic density in the inner core increases, triggering the nucleonic and hyperonic DU processes as shown in Figs.~\ref{fig:PolarPlot}-\ref{fig:PolarPlot2}. Thus, the onset of the enhanced neutrino emission occurs at the same particle fractions and central baryonic density for stars with and without DM. However, as it can be seen in Fig.~\ref{fig:DU_ParticleFraction} and Table~\ref{Table:DUrcas_IST_FSU2R}, with the increase of the DM fraction the total gravitational mass at which the DU processes are kinematically allowed is shifted towards lower masses. Thus, for the FSU2H EoS the mass of the star at the nucleonic DU onset, $M_{DU}$, drops from 1.85 M$_{\odot}$ to 1.75 M$_{\odot}$, 1.71 M$_{\odot}$, and 1.66 M$_{\odot}$ for 2\%, 3\%, and 4\% of DM, respectively (see Table~\ref{Table:DUrcas_FSU2H}). Moreover, due to hyperonic DU processes, such as $\Lambda p$ and $\Sigma^- \Lambda$, triggered before the nucleonic DU process, $np$, an accumulation of DM allows for an NS to exhibit a fast cooling behavior at 1.24 M$_\odot$, for 4\% DM (as shown in Fig.~\ref{fig:DU_ParticleFraction}). The values for the central baryonic densities and total gravitational masses at which the hyperonic and nucleonic DU processes become active are presented in Table~\ref{Table:DUrcas_FSU2H}.

As was discussed in Ref.~\cite{Avila:2023rzj}, another consequence of the presence of the DM core is related to the reduction of the stellar radius due to the stronger gravitational pull compared to a pure BM configuration. This effect is presented in Table~\ref{Table:radii}. As the photon emission is related to the stellar radius, this effect alters the photon luminosity that becomes important for stars of $\gtrsim 10^{5}$ years~\citep{Yakovlev:2004iq}.

\begin{table}[ht]
\centering
\resizebox{0.5\columnwidth}{!}{
\begin{tabular}{|c|cccc|}
\hline
\multirow{0}{*}{IST EoS}   & \multicolumn{4}{c|}{$\mathrm{f_{DM}}$}                                                                            \\ \cline{2-5} 
                           & \multicolumn{1}{c|}{$0\%$} & \multicolumn{1}{c|}{$2\%$} & \multicolumn{1}{c|}{$3\%$} & $4\%$                      \\ \hline
$M_\mathrm{tot}$ [$\mathrm{M_\odot}$]     & \multicolumn{4}{c|}{$R_\mathrm{B}$ [km]}                                                                                       \\ \hline
1.20                       & \multicolumn{1}{c|}{11.29} & \multicolumn{1}{c|}{11.11} & \multicolumn{1}{c|}{11.03} & 10.94                      \\ \hline
1.60                       & \multicolumn{1}{c|}{11.10} & \multicolumn{1}{c|}{10.91} & \multicolumn{1}{c|}{10.81} & 10.70                      \\ \hline
1.90                       & \multicolumn{1}{c|}{10.58} & \multicolumn{1}{c|}{10.35} & \multicolumn{1}{c|}{10.20} & 10.05                      \\ \hline \hline
\multirow{0}{*}{FSU2R EoS} & \multicolumn{4}{c|}{$\mathrm{f_{DM}}$}                                                                            \\ \cline{2-5} 
                           & \multicolumn{1}{c|}{$0\%$} & \multicolumn{1}{c|}{$2\%$} & \multicolumn{1}{c|}{$3\%$} & $4\%$                      \\ \hline
$M_\mathrm{tot}$ [$\mathrm{M_\odot}$]     & \multicolumn{4}{c|}{$R_\mathrm{B}$ [km]}                                                                                       \\ \hline
1.20                       & \multicolumn{1}{c|}{12.18} & \multicolumn{1}{c|}{12.09} & \multicolumn{1}{c|}{12.01} & 11.93                      \\ \hline
1.60                       & \multicolumn{1}{c|}{12.39} & \multicolumn{1}{c|}{12.25} & \multicolumn{1}{c|}{12.15} & 12.05                      \\ \hline
1.90                       & \multicolumn{1}{c|}{12.16} & \multicolumn{1}{c|}{11.93} & \multicolumn{1}{c|}{11.73} & \multicolumn{1}{c|}{11.47} \\ \hline \hline
\multirow{0}{*}{FSU2H EoS} & \multicolumn{4}{c|}{$\mathrm{f_{DM}}$}                                                                            \\ \cline{2-5} 
                           & \multicolumn{1}{c|}{$0\%$} & \multicolumn{1}{c|}{$2\%$} & \multicolumn{1}{c|}{$3\%$} & $4\%$                      \\ \hline
$M_\mathrm{tot}$ [$\mathrm{M_\odot}$]     & \multicolumn{4}{c|}{$R_\mathrm{B}$ [km]}                                                                                       \\ \hline
1.20                       & \multicolumn{1}{c|}{12.43} & \multicolumn{1}{c|}{12.30} & \multicolumn{1}{c|}{12.24} & 12.16                      \\ \hline
1.50                       & \multicolumn{1}{c|}{12.77} & \multicolumn{1}{c|}{12.62} & \multicolumn{1}{c|}{12.53} & 12.44                     \\ \hline
1.70                       & \multicolumn{1}{c|}{12.86} & \multicolumn{1}{c|}{12.63} & \multicolumn{1}{c|}{12.50} & \multicolumn{1}{l|}{12.38} \\ \hline
\end{tabular}
}
\caption{Baryonic radii for different gravitational mass configurations for the three models used in this work and their modifications due to the addition of DM.}
\label{Table:radii}
\end{table}


\subsection{Mapping the nucleonic and hyperonic DU regions inside the star}
\label{subsec:pieplots}

\begin{figure}[t]
    \centering
    \includegraphics[scale=0.5]{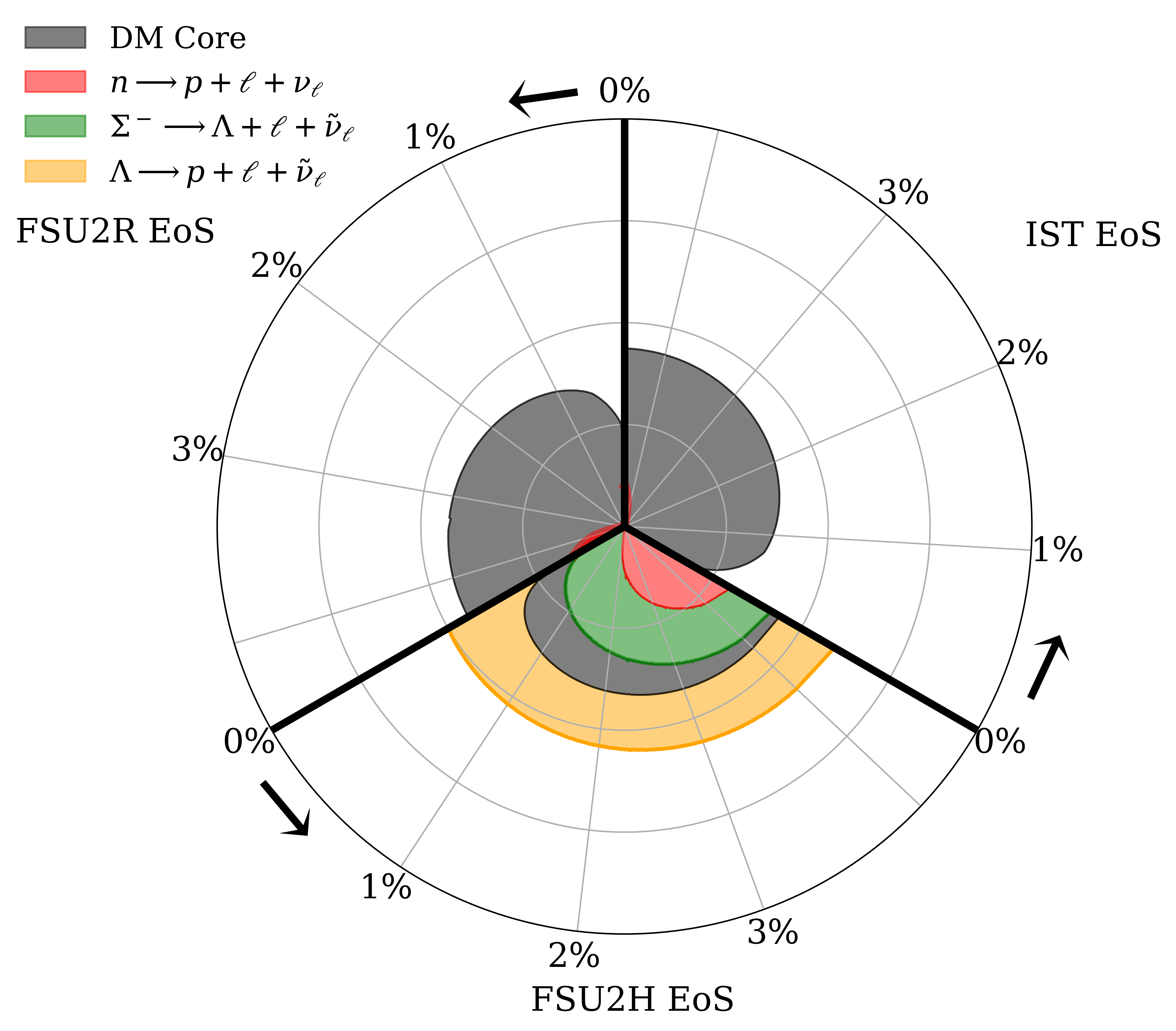}
    \caption{Stellar configurations with different DM fractions for the IST EoS (top right slice), FSU2R EoS (top left slice), and FSU2H EoS (bottom slice). The size of the $np$, $\Sigma^-\Lambda$, $\Lambda p$ DU regions, and DM core are depicted in light red, light green, yellow, and dark gray, respectively. For a better comparison, the radii are normalized to the baryonic radius of each configuration and are given in Table~\ref{Table:radii}. All configurations correspond to NSs with a total gravitational mass of 1.75 M$_\odot$.}
    \label{fig:PolarPlot}
\end{figure}

To investigate an interplay between the size of the DM core and the star regions in which $n p$, $\Lambda p$ and $\Sigma^-\Lambda$ DU processes are operating we show the pie charts, where each slice represents a different BM model. Thus, in Fig.~\ref{fig:PolarPlot} the top right, top left and bottom slices illustrate stars of 1.75 M$_\odot$ described by the IST, FSU2R, and FSU2H EoSs, respectively. The DM fraction continuously changes from 0$\%$ to 4.5$\%$. All radii of the DM-admixed configurations are normalized to the baryonic radius of each configuration. The physical values of radii and total gravitational masses are listed in Table~\ref{Table:radii}.

As it is shown in Tables~\ref{Table:DUrcas_IST_FSU2R},~\ref{Table:DUrcas_FSU2H} none of the pure BM stars of 1.75 M$_\odot$ have the operating nucleonic DU in their interior. However, by increasing the DM fraction the $n p$ DU process is triggered at 3.87$\%$, 4.17$\%$, 1.96$\%$ for the IST, FSU2R, and FSU2H EoSs respectively. A much lower value for the FSU2H EoS is related to the fact that the nucleonic DU is triggered at a lower stellar mass, i.e. $1.85$ M$_\odot$, than for the IST and FSU2R EoSs, i.e. $1.908$ M$_\odot$ and $1.921$ M$_\odot$ (see Tables~\ref{Table:DUrcas_IST_FSU2R},~\ref{Table:DUrcas_FSU2H}). The nucleonic DU region is depicted with the light red color in Fig.~\ref{fig:PolarPlot}. Moreover, parts of the star in which the $\Lambda p$ and $\Sigma^-\Lambda$ DU processes take place are shown in yellow and light green colors. 

\begin{figure}[t]
    \centering
    \includegraphics[scale=0.5]{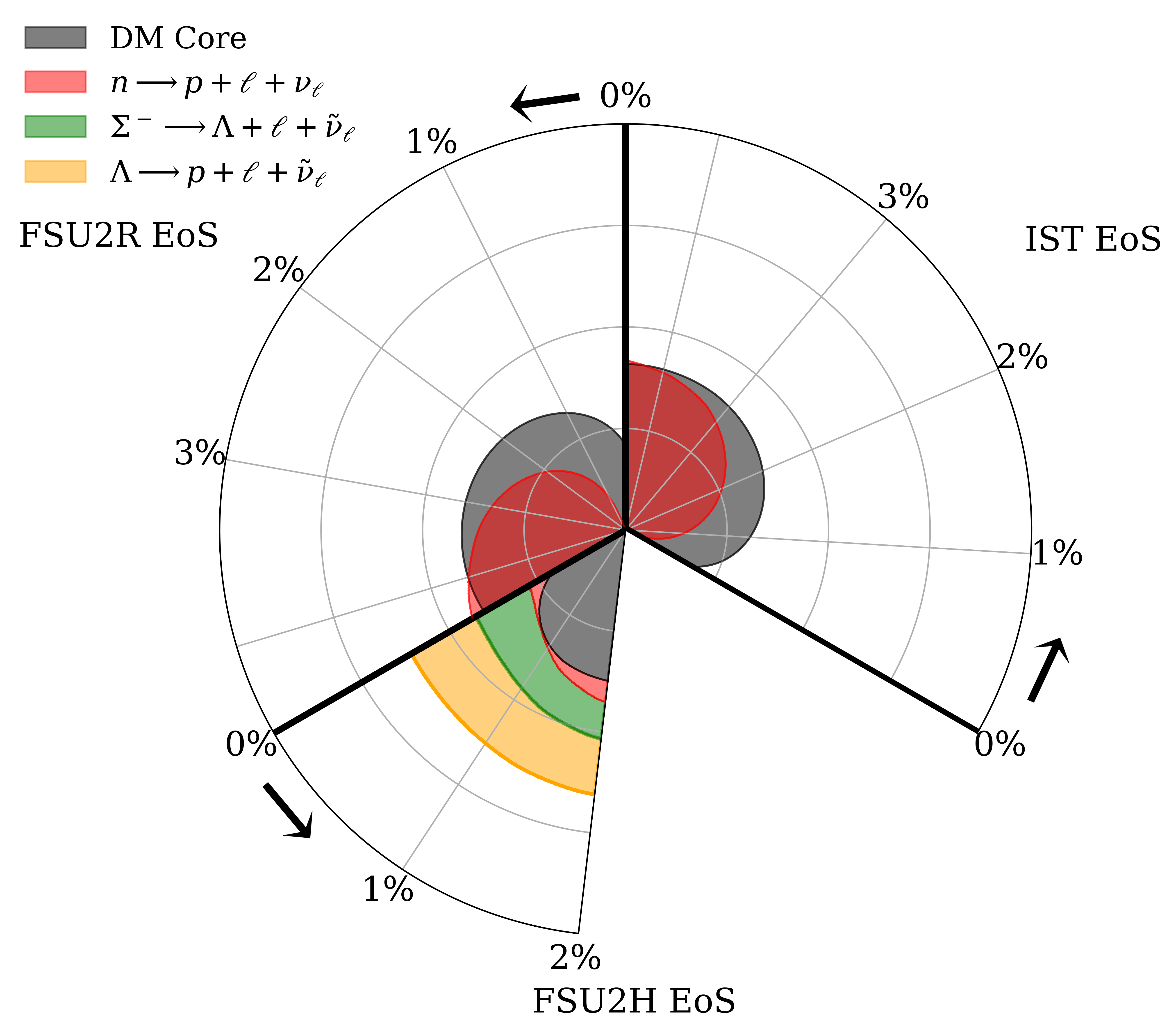}
    \caption{The same as in Fig.~\ref{fig:PolarPlot}, but for the total gravitational mass of 1.9 M$_\odot$. The empty part of the bottom slice corresponds to the non-existing configurations (see details in the text).
    }
    \label{fig:PolarPlot2}
\end{figure}

For a comparison, Fig.~\ref{fig:PolarPlot2} shows stars of $1.90$ M$_\odot$ described with the same models as in Fig.~\ref{fig:PolarPlot}. In the case of the FSU2H EoS, not only the hyperonic DU processes but also the nucleonic DU are activated. Moreover, as it can be seen in Fig.~\ref{fig:PolarPlot2}, no DM-admixed configurations with a DM fraction higher than 2\% exist in this model. This is related to the FSU2H EoS that has an upper limit on the baryonic density, $n_B=1$ fm$^{-3}$, at which the effective nucleonic mass vanishes~\citep{Tolos:2017lgv}. As higher DM fractions require higher central baryonic densities, which can not be obtained within the FSU2H EoS, such configurations do not exist. 

\subsection{Cooling curves}

\begin{figure}[t]
\centering
\includegraphics[width=0.48\linewidth]{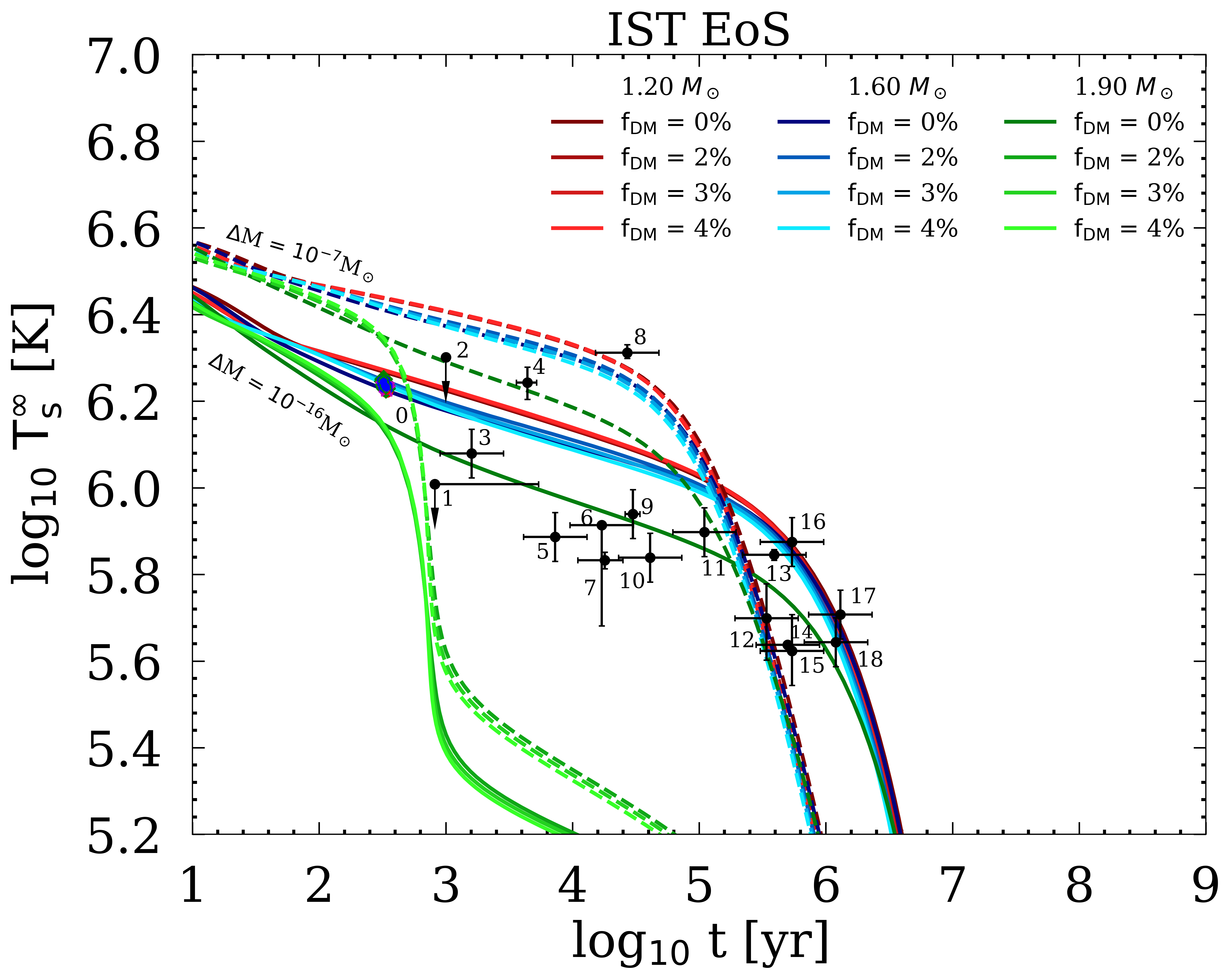}
\includegraphics[width=0.48\linewidth]{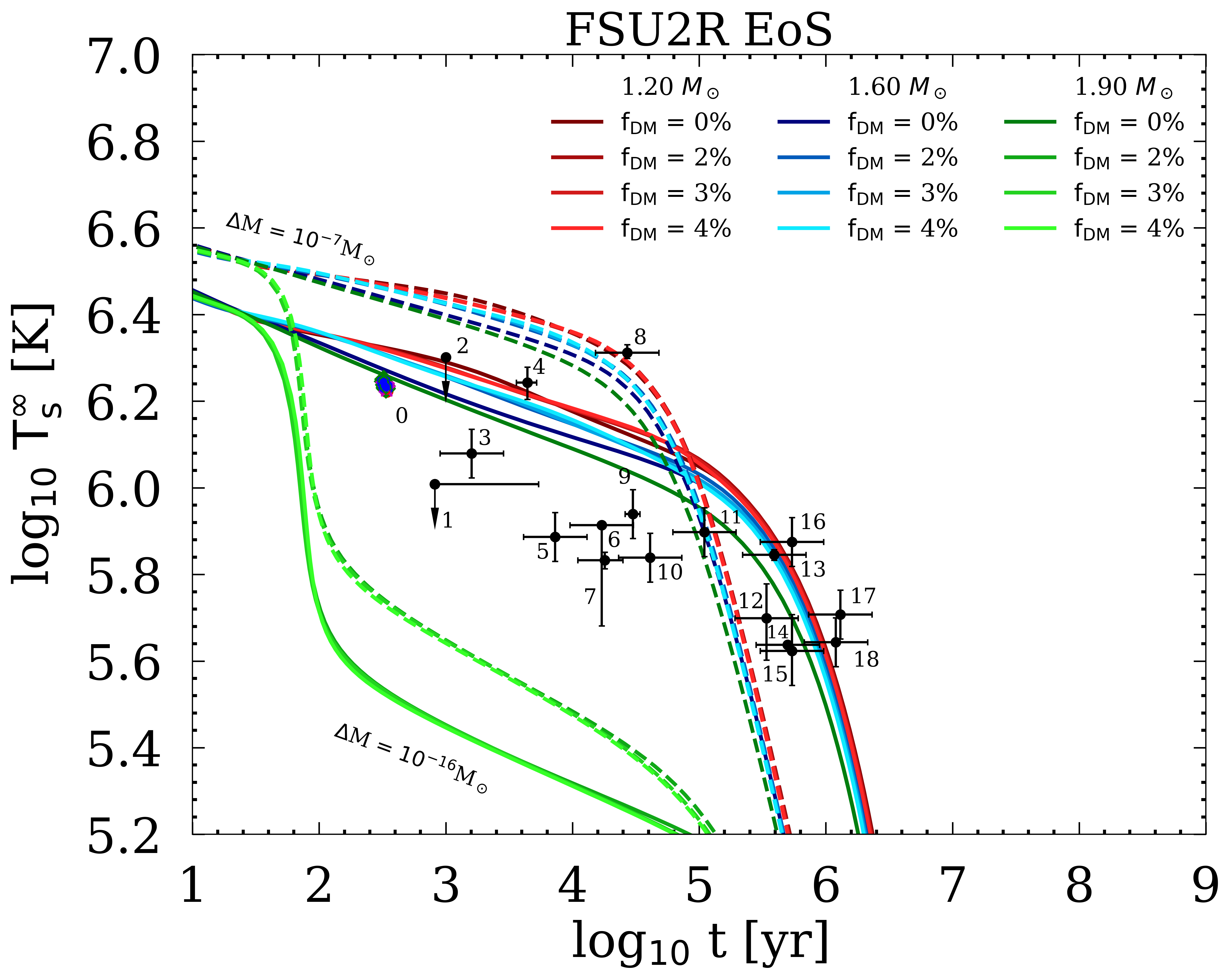}
\caption{Cooling curves for stars with total gravitational masses of $M = 1.2, 1.6$, and $1.9$ M$_\odot$ described by the IST EoS (left) and FSU2R EoS (right) are shown. The considered DM fraction is $f_{DM}=0\%,\ 2\%,\ 3\%$ and $4\%$ for particle's mass $m_{DM} = 1$ GeV. The solid and dashed curves correspond to envelopes composed of heavy elements ($\eta = \Delta M/M = 10^{-16}$) and light elements ($\eta = \Delta M/M = 10^{-7}$), respectively. The impact of neutron superfluidity in the $^{1}S_0$ channel, employing the SFB model~\cite{Schwenk:2002fq}, and proton superconductivity in the $^{1}S_0$ channel, employing the CCDK model~\cite{Chen:1993bam}, is considered. The figure is adopted from~\cite{Avila:2023rzj}. The utilized observational data are listed in the Appendix.}
\label{fig:CoolingISTFSU2RDM}
\end{figure}

Fig.~\ref{fig:CoolingISTFSU2RDM} depicts the red-shifted surface temperature as a function of age for stars of 1.2 M$_\odot$ (red curves), 1.6 M$_\odot$ (blue curves), and 1.9 M$_\odot$ (green curves) modeled within the IST EoS (left panel) and the FSU2R EoS (right panel). The cooling curves for the stars of 1.2 M$_\odot$ (red curves), 1.5 M$_\odot$ (blue curves), and 1.7 M$_\odot$ (green curves) modeled within the FSU2H EoS are shown in Fig.~\ref{fig:CoolingFSU2HDM}. This choice of lower mass values in comparison to Fig.~\ref{fig:CoolingISTFSU2RDM} is related to the upper limit of the baryonic density and absence of heavy stars admixed with DM within the FSU2H model. 

The cooling curves were obtained considering the neutron and proton $^1S_0$ pairing, described by the SFB~\cite{Schwenk:2002fq} and CCDK~\cite{Chen:1993bam} models, as the ones that provide the best description of the observational data. Following~\cite{Avila:2023rzj}, in Figs.~\ref{fig:CoolingISTFSU2RDM},~\ref{fig:CoolingFSU2HDM} the color grade represents the different DM fractions, whereas the higher DM fraction corresponds to a lighter shade. Pure BM stars, i.e. with 0$\%$ of DM, are shown with the darkest shade of each color.

The envelope composition plays an important role in modeling the surface luminosity as it affects the relation between the surface and core temperatures. In particular, the relation between the internal and surface temperature is very sensitive to the outermost layer composition, which plays the role of an insulator until very low temperatures. To address this question, we consider the fraction of light elements via the factor $\eta=\Delta M/M$ with $\Delta M$ being the mass of light elements in the envelope. In Figs.~\ref{fig:CoolingISTFSU2RDM},~\ref{fig:CoolingFSU2HDM} the light-element envelope, i.e. hydrogen, helium, with $\eta=\Delta M/M= 10^{-7}$ is depicted with the dashed curves, while the heavy-elements envelope, mostly carbon, with $\eta=\Delta M/M= 10^{-16}$, is shown with solid curves. 

On both panels in Fig.~\ref{fig:CoolingISTFSU2RDM} the cooling curves for the 1.9${M_\odot}$ stars with 0$\%$ of DM show a slow cooling, corresponding to the absence of the DU process. However, an accrued DM of $m_{DM} = 1$ GeV of $f_{DM}\simeq0.161\%$ (IST EoS) and $f_{DM}=0.378\%$ (FSU2R EoS) triggers the previously forbidden process to operate. Thus, the green curves for 
$f_{DM}=2\%,\ 3\%$ and $4\%$ exhibit a rapid drop of the surface temperature.

As seen in Fig.~\ref{fig:CoolingFSU2HDM} the FSU2H model provides a good agreement with the observational data that fully agrees with the results of Ref.~\cite{Negreiros:2018cho}. For the pure BM star of 1.7${M_\odot}$ only the $\Lambda p$ DU process operates leading to a medium cooling. At $f_{DM}\simeq0.61\%$ the $\Sigma^- \Lambda$ DU process onsets causing a rapid temperature drop, while the $np$ DU process takes place for much higher DM fractions (see light green curves in Fig.~\ref{fig:CoolingFSU2HDM}).

\subsection{Thermal evolution of Cassiopeia A as a DM-admixed NS}

The central compact object (CCO) in Cassiopeia A (Cas A) has been raising various discussions due to its unusually rapid cooling~\cite{Ho:2009mm,Heinke:2010cr,Shternin:2010qi,Elshamouty:2013nfa,Wijngaarden:2019tht}. Recent combined analysis of the X-ray spectra of Cas A suggests that the surface temperature drop is $2.2\pm 0.3$ \% over 10 years when the absorbing hydrogen column density $\mathrm{N_H}$ is used as a free model parameter, and $1.6 \pm 0.2$ \% for a constant $\mathrm{N_H}$ value~\citep{Shternin:2022rti}. The thermal evolution of such a young NS, with an estimated age around 356 years, opened up a pandora box of models, between which the $^3P_2$ pairing of neutrons in the core~\citep{Page:2010aw,Ho:2014pta}, rapid cooling via the DU process~\citep{Taranto:2015ubs}, the impact of medium-modified one-pion exchange in dense matter~\citep{Blaschke:2011gc} and beyond the Standard Model physics~\citep{Hamaguchi:2018oqw}. As the estimated mass of the CCO in Cas A is around $1.55$ M$_\odot$~\cite{Shternin:2022rti}, a realistic description should take place for a middle mass star. 

As it was shown in Ref.~\cite{Tsiopelas:2020nzm}, the IST EoS does not necessarily require the incorporation of the neutron superfluidity and/or proton conductivity to explain the Cas A temperature drop. The model reproduced the Cas A data with 1.66 M$_\odot$ and 1.91 M$_\odot$ stars with the inclusion of neutron and proton singlet pairings, as well as with the 1.96 M$_\odot$ star for unpaired matter~\cite{Tsiopelas:2021ayd}. 

As it was shown in~\cite{Avila:2023rzj} the FSU2R EoS describes the Cas A cooling with a combination of n $^{1}S_0$ (SFB model), p $^{1}S_0$ (CCDK model), n $^{3}P_2$ pairing (T72 model) with the maximum critical temperature  $T_{c} = 7.105 \cdot 10^{8}$ K. 
In Ref.~\cite{Negreiros:2018cho}, the best fit of Cas A within the FSU2H EoS is obtained for the $1.88$ M$_\odot$ star considering the proton and neutron single pairing together with the triplet neutron pairing characterized by the maximum critical temperature $T_{c} \sim 1.41 \cdot 10^{9}$ K. For the same EoS, we also see a better agreement with Cas A data while varying the neutron $^3P_2$ pairing.

\begin{figure}[t]
    \centering
    \includegraphics[width=0.7\linewidth]{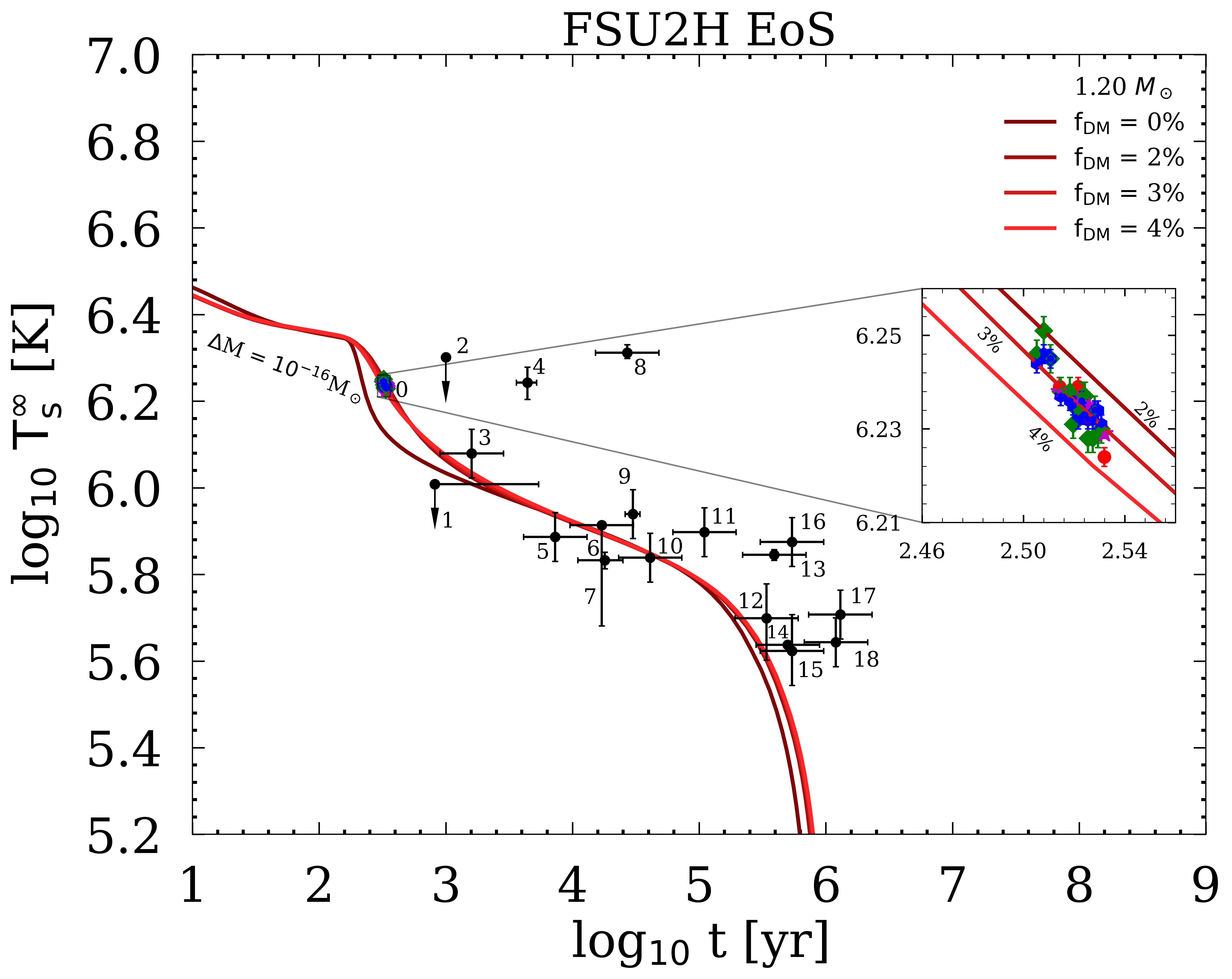}
    \caption{The results of the fit of the surface temperature as a function of age for the CCO in Cas A measured by Chandra ACIS-S in GRADED and FAINT modes. The red and magenta data points correspond to variable and fixed absorbing hydrogen column density $\mathrm{N_H}=1.656 \cdot 10^{22}~ cm^{-2}$ in the FAINT mode, while the green and blue data points depict the same data, but in the GRADED mode. 
    }
    \label{fig:Cas}
\end{figure}

As one of the results of our analysis, the best fit is obtained for the DM-admixed star of $1.2$ M$_\odot$ with a DM fraction of $3\%$. The results of the fit are presented in Fig.~\ref{fig:Cas} whereas the curves are obtained for the heavy-elements envelope $\eta=\Delta M/M= 10^{-16}$ and a combination of n$^1S_0$ (SFB model~\cite{Schwenk:2002fq}), p$^1S_0$ (CCDK model~\cite{Chen:1993bam}), and n$^3P_2$ pairing (T72 model~\citep{10.1143/PTP.48.1517}) with the maximum critical temperature $T_{c} \sim 8.52 \cdot 10^{8}$ K.

\section{Conclusions}
\label{sec:Concl}

Compact star cooling is an important probe of the particle composition and physics governing the internal dynamics. In this work, we extend a study of~\citet{Avila:2023rzj} of the impact of asymmetric fermionic DM on the thermal evolution of NSs. Within this scenario, DM does not self-annihilate due to the particle-antiparticle asymmetry and, therefore, is accrued in the star's interior. We demonstrate that considering its interaction with BM through gravity the accumulation of DM leads not only to the well-known modification of the star's gravitational mass, radius, and tidal deformability but also affects its thermal evolution. Thus, accrued DM indirectly alters the cooling behavior of NSs by pulling inward BM from the outer layers, increasing the central density and consequently affecting the BM distribution. The latter triggers the onset of nucleonic and hyperonic DU processes in lower NS masses and rapid star cooling due to the enhanced neutrino emission. At the same time, the proton fraction corresponding to the nucleonic DU onset remains equivalent to the one of a pure BM star. The same occurs with the onset density of hyperons. These onset densities 
correspond to the NSs of smaller masses due to their compactification. 

\begin{figure}[t]
\centering
\includegraphics[width=0.7\linewidth]{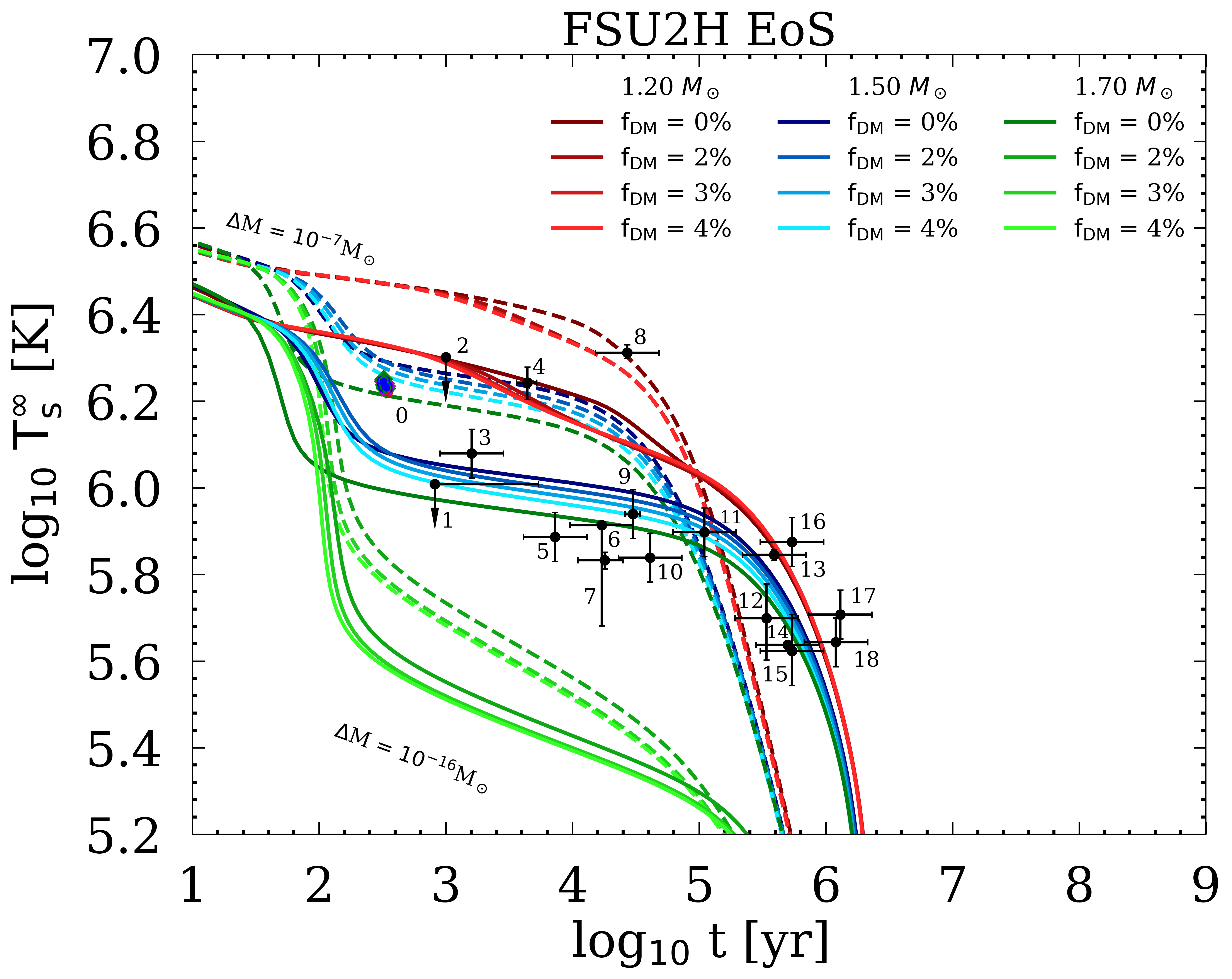}
\caption{The same as in Fig.~\ref{fig:CoolingISTFSU2RDM}, but for stars with the total gravitational masses of $M = 1.2, 1.5, 1.7$ M$_\odot$ modeled within the FSU2H EoS and supplemented with n$^3P_2$ pairing described by the T72 model.}
\label{fig:CoolingFSU2HDM}
\end{figure}

Another consequence of the gravitational pull of BM towards the center results in a reduction of the baryonic radius of the star. Therefore, the stellar surface decreases, leading to a lower photon luminosity. This effect becomes visible during the photon-dominated stage when neutrino emission plays a secondary role.

To thoroughly study this effect, we considered the nucleonic IST and FSU2R models that have the DU onset at 1.908 M$_{\odot}$ and 1.921 M$_{\odot}$, respectively, and the hyperonic FSU2H model that kinematically allows both nucleonic and hyperonic DU processes. Accumulation of the DM particles with a mass of $m_\chi = 1$ GeV of $f_\chi\simeq0.161\%$ (IST EoS) and $f_\chi=0.378\%$ (FSU2R EoS) triggers the previously energetically forbidden process. 

At the same time, the FSU2H EoS gives the possibility to study the sequence of the DU onsets with an increase of the DM fraction, both due to nucleonic and hyperonic DU processes. Thus, for the 1.7${M_\odot}$ star with $0\%$ of DM, only the $\Lambda p$ DU process operates leading to a medium cooling. At $f_{DM}\simeq0.61\%$ the $\Sigma^- \Lambda$ DU process onsets causing a rapid temperature drop, while the final contribution of $np$ DU process occurs for $f_{DM}> 2\%$.

We show that the rapid cooling of the CCO in Cas A could be explained by the DM-admixed star of $1.2$ M$_\odot$ described by the FSU2H EoS. In~\cite{Avila:2023rzj} the best fit of the Cas A cooling rate was obtained for the $M =1.6~{\rm M}_\odot$ star modeled within the FSU2R EoS with $f_{DM}=4\%$ and light-elements envelope. Thus, the accrued DM helps to reconcile the star mass at which the DU process is kinematically allowed with the observational data on the mass of the CCO in Cas A by lowering the mass compared to a pure BM star. We demonstrate that an increase of the DM fraction causes a shift of the DU onset towards lower gravitational masses of the star. This effect could serve as a distinctive signature of the presence of DM in compact stars. In Ref.~\cite{Avila:2023rzj} it was shown that a similar result can be obtained by considering heavier DM particles, leaving the DM fraction to be low. 

Thus, low/middle mass NSs that are not expected to have the operated DU process, might show a rapid cooling due to the presence of DM. Consequently, stars of similar masses will display different cooling patterns depending on the accrued DM fraction in their interiors. As DM fraction increases towards the Galactic center, the thermal evolution of NSs might exhibit distinct features compared to stars further away, e.g. in the Solar vicinity and beyond. However, the measurements of the distant NSs are not so precise. Therefore, several radio telescopes such as FAST~\cite{Nan:2011um}, SKA~\cite{Carilli:2004nx}, and CHIME~\cite{CHIMEPulsar:2020uip} will search for the old, isolated compact stars near the Earth. The recently launched JWST together with the forthcoming Thirty Meter Telescope (TMT) and Extremely Large Telescope (ELT)~\cite{TMTInternationalScienceDevelopmentTeamsTMTScienceAdvisoryCommittee:2015pvw} will provide a sensitive probe of old NSs~\cite{Chatterjee:2022dhp}. 
These searches together with studies of the NS matter EoS are important for understanding the interior composition of compact stars and possible degeneracy between the effects of strongly interacting matter and DM~\cite{Giangrandi:2022wht}.

\appendix

\section*{Appendix: Observational data}
\label{sec:data}

Observational data of the CCO in the Cas A supernova remnant are depicted as source 0 in Figs.~\ref{fig:CoolingISTFSU2RDM},~\ref{fig:CoolingFSU2HDM},~\ref{fig:Cas}. It shows the temperature measurements using Chandra ACIS in GRADED and FAINT modes with 1$\sigma$ error bars~\cite{Ho:2014pta}. The surface temperature is subject to variation across different observations and the hydrogen column density $\mathrm{N_H}$ may also vary~\citep{Shternin:2022rti}, leading to different $T{_s^\infty}$ values. Thus, the plotted data are the outcome of examining both the GRADED and FAINT modes mode of the ACIS detector. One data set involves varying the hydrogen column density, while the other considers it fixed.

The rest of the data shown in Figs.~\ref{fig:CoolingISTFSU2RDM},~\ref{fig:CoolingFSU2HDM},~\ref{fig:Cas} were taken from Ref.~\cite{Beznogov:2014yia}.  We consider 2$\sigma$ error bars for the available data, otherwise the factor of 0.5 and 2 in the temperature and age, excluding the upper limits, is utilized. The sources are: 1 - PSR J0205+6449 (in 3C58), 2 - PSR B0531+21 (Crab), 3 - PSR J1119-6127, 4 - RX J0822-4300 (in PupA), 5 - PSR J1357-6429, 6 - PSR B1706-44, 7 - PSR B0833-45 (Vela), 9 - PSR J0538+2817, 10 - PSR B2334+61, 11 - PSR B0656+14, 12 - PSR B0633+1748 (Geminga), 13 - PSR J1741-2054, 14 - RX J1856.4-3754, 15 - PSR J0357+3205 (Morla), 16 - PSR B1055-52, 17 - PSR J2043+2740, 18 - RX J0720.4-3125. The surface temperature of the object 8 - XMMU J1731-347 from Ref.~\cite{Beznogov:2014yia} was substituted by the updated results HESS J1731-347 from~\cite{Doroshenko2022}, while the age of the object is considered from Ref.~\cite{Beznogov:2014yia}.

\vspace{6pt} 

\authorcontributions{E.G. and A.Á. carried out simulations, data analysis, and preparation of figures. V.S., O.I., and C.P. proposed the idea and contributed to the data interpretation. All authors have contributed to the preparation of the manuscript and discussions. All authors have read and agreed to the published version of the manuscript.}

\funding{The work is supported by the FCT – Fundação para a Ciência e a Tecnologia, within the project No. EXPL/FIS-AST/0735/2021 with DOI identifier 10.54499/EXPL/FIS-AST/0735/2021. A.Á., E.G., V.S., and C.P. acknowledge the support from FCT – Fundação para a Ciência e a Tecnologia within the projects UIDP/\-04564/\-2020 and UIDB/\-04564/\-2020, respectively, with DOI identifiers 10.54499/UIDP/04564/2020 and 10.54499/UIDB/04564/2020. E.G. also acknowledges the support from Project No. PRT/BD/152267/2021. C.P. is also supported by project No. 2022.06460.PTDC with the  DOI identifier 10.54499/2022.06460.PTDC. The work of O.I. was supported by the program Excellence Initiative--Research University of the University of Wrocław of the Ministry of Education and Science.}
\institutionalreview{Not applicable.}

\informedconsent{Not applicable.}

\dataavailability{Not applicable.} 


\conflictsofinterest{The authors declare no conflict of interest.} 

\sampleavailability{Not applicable.}


\abbreviations{Abbreviations}{
The following abbreviations are used in this manuscript:\\

\noindent 
\begin{tabular}{@{}ll}
EoS   & equation of state \\
NS    & neutron star \\
BM    & baryonic matter \\
DM    & dark matter \\
DU    & direct Urca \\
MU    & modified Urca \\
GW    & gravitational wave \\
PBF   & pair breaking and formation \\
IST   & Induced Surface Tension \\
TOV   & Tolmann-Oppenheimer-Volkoff\\
JWST  & James Webb Space Telescope\\

\end{tabular}
}

\begin{adjustwidth}{-\extralength}{0cm}

\reftitle{References}
\bibliography{bibliography}

\PublishersNote{}
\end{adjustwidth}
\end{document}